\documentclass[]{jfm}

\usepackage{graphicx}
\usepackage{newtxtext}
\usepackage{newtxmath}
\usepackage{natbib}
\usepackage{hyperref}
\usepackage[capitalise]{cleveref}
\usepackage{subcaption}
\definecolor{mygray}{gray}{0.2}
\hypersetup{
    colorlinks = true,
    urlcolor   = blue,
    citecolor  = black,
}

\newcommand{\RomanNumeralCaps}[1]

\title{Direct numerical simulation of particle clustering and turbulence modulation: an Eulerian approach}

\author{Ajay Dhankarghare\aff{1},
 \and Yuval Dagan\aff{1}\corresp{\email{yuvalda@technion.ac.il}}}

\affiliation{\aff{1}Faculty of Aerospace Engineering, Technion - Israel Institute of Technology, Haifa, 3200003, Israel}

\begin{document}
\maketitle
\begin{abstract}
We present a new Eulerian framework for the computation of turbulent compressible
multiphase channel flows, specifically to assess turbulence modulation by dispersed
particulate matter in dilute concentrations but with significant mass loadings. By
combining a modified low-dissipation numerical scheme for the carrier gas phase and
a quadrature-based moment method for the solid particle phase, turbulent statistics of the
fluid phase and fluctuations of the particle phase may be obtained as both are resolved
as coupled fields. Using direct numerical simulations, we demonstrate how this method
effectively resolves the turbulent statistics, kinetic energy, skin friction drag, particle mass
flow rate and interphase drag for moderate-Reynolds-number channel flows for the first
time. Validation of our approach to the turbulent particle-free flow and the turbulent
particle-laden flow proves the applicability of the carrier flow low-dissipation scheme
to simulate relatively low-Mach-number compressible flows and of the quadrature-based
moment method to simulate the particle phase as an Eulerian field. This study also
rationalises the computed interphase drag modulation and total Reynolds shear stress
results using a simplified analytical approach, revealing how the particle migration towards
the wall can affect the drag between the two phases at different Stokes numbers and
particle loadings. Furthermore, we show the effect of near-wall particle accumulation on
the particle mass flow rate. Using our Eulerian approach, we also explore the complex
interplay between the particles and turbulent fluctuations by capturing the preferential
clustering of particles in turbulence streaks. This interplay leads to turbulence modulations
similar to recent observations reported in prior computational works using Lagrangian
simulations. Our study extends the applicability of the Eulerian approach to accurately
study particle–fluid interactions in compressible turbulent flows by explicitly calculating
the energy equations for both the particle phase and the carrier fluid motion. Since the
formulation is compressible and includes energy equations for both the particle and carrier flow fields, future studies for compressible flows involving heat and mass transfer may be
simulated using this methodology.
\end{abstract}

\maketitle

\section{\label{sec:intro}Introduction}
The analysis of dispersed particles in multiphase flow using high-fidelity simulations may
require the implementation of Lagrangian particle tracking \citep{Balachandar2010}.
Using simplified theoretical flow models, one may employ Lagrangian methods to track
the particle and droplet trajectories and study their dispersion \citep{Marcu1996I,Marcu1996mixing,Marcu1996II} and entrapment within coherent flow structures \citep{Marcu1995burgers, IJzermans2006, Angilella2010, dagan2021settling, Avni2022a, Ravichandran2022, avni2023dropleta, avni2023dropletb}. Following
such a simplified approach allows the isolation of specific transport mechanisms, including
oscillatory flows \citep{Dagan2017b,dagan2017particle,daganSimilarityFlames2018}, aerosol formation \citep{Avni2022b}
and particle structures \citep{Yerasi2022}, while studying their
influence on particle dynamics. However, incorporating detailed models for such cases
might result in computationally intensive modelling, even for relatively simple set-ups.
The approach solves the equation for each particle, which may restrict its use when dealing
with an extremely large number of particles due to high computation costs. Furthermore,
simulating the particles with low Stokes numbers adds to the costs, as low-inertia particles
may impose strict limitations on the maximum numerical time step. \cite{salman2004lagrangian} demonstrated that the Lagrangian procedure might not converge under mesh
refinement, which can be problematic when the mesh used is non-uniform, when the mesh
size is less than the particles’ size or when polydisperse particles are used. In the case
of a very sparsely populated domain, the method can lead to self-induced motion of the
particles due to their feedback force on the flow. Many intriguing phenomena that result
from particle–fluid interactions in turbulent channel flows, such as turbophoresis, preferential clustering, uneven drag and turbulence modulation, have been studied using the
Lagrangian approach. However, statistical study of these phenomena requires resolving the
fluctuations in particle properties, such as particle number density and particle velocities
within a control volume, that might be difficult to estimate using the Lagrangian approach.

The Eulerian approach to simulating the particle phase offers a more efficient and
significantly less computationally expensive alternative. Earlier works \citep{zhang1994averaged,druzhinin1994concentration,druzhinin1995two,zhang1997momentum,mehta1998fast}
used the Eulerian approach to simulate the dispersed-particle field. The approach assumes
an average particle field inside the control volume, allowing for the simulation of particles
in the domain with a non-uniform mesh that can be smaller than the particle size. However,
averaging the particle field limits the ability of the approach to resolve various particle
phenomena like particle-trajectory crossing, particle–wall reflections and inter-particle
collisions. Furthermore, the absence of a pressure term in the particle phase transport
system leads to a non-hyperbolic nature of the system, which can result in numerical
stiffness problems \citep{bouchut1994zero}. To overcome these limitations, Eulerian approaches
were developed using higher-moment methods \citep{desjardins2008quadrature,fox2008numerical,Marchisio_Fox_2013,vie2015anisotropic,patel2017verification,kasbaoui2019clustering,schneiderbauer2019numerical,heylmun2021quadrature}. Studies using these methods are mostly restricted to the investigation
of particle-laden flows in isotropic homogeneous turbulence, where wall reflections can be avoided. \cite{baker2020verification} used the Eulerian moment method approach to simulate
the particle field in a vertical-channel turbulent fluid flow and compared the results with
those using the Lagrangian approach. They noticed that simulations of the flow with low
Stokes numbers gave similar results for the two approaches. At high Stokes number and
low mass loading, the Eulerian approach failed to predict similar results to the Lagrangian
approach due to its inability to simulate a very sparsely populated particle field. The
authors considered the anisotropic Gaussian distribution of the particle velocity \citep{kong2017euler}. \cite{fox2012large} and \cite{baker2020verification} claimed that using this distribution
with up to second-order moments of velocity can capture the anisotropic velocity of
the particle phase but not the particle-trajectory crossing for which velocity moments
of up to third order were required. Though the authors showcased the effectiveness of
the moment method in simulating the particle field with a significant mass loading in
low Stokes flow, the ability of the method to simulate the particle field in different
environments and configurations remains elusive. It is also required for the method to
be able to capture the particle-trajectory crossing. This can be crucial for future studies of
a very large number of particles in both compressible and incompressible flows involving
various important phenomena such as heat transfer, coalescence, inter-particle collisions,
combustion, evaporation, condensation and dispersion. These phenomena can significantly
affect the stability of turbulent flows \citep{dagan2015dynamics,dagan2016evolution}. Furthermore, it is necessary
to establish the reliability of the method to capture various intriguing phenomena that
result from particle–fluid interactions in a turbulent channel flow, such as uneven drag,
turbophoresis, preferential clustering and turbulence modulation.

The primary objective of this study is to introduce an Eulerian approach for the
simulation of particle fields in channel flows. The goal is to demonstrate the efficacy of this
approach in accurately predicting particle behaviour in turbulent flows, and particularly,
to resolve the particle–fluid interactions affecting turbulence modulation, drag, particle
accumulation and preferential clustering. Moreover, the capability of a compressible
solver to effectively handle the flow field even in an incompressible regime ($M<0.3$) is
explored.

Particles in turbulent flows tend to migrate towards the wall by a phenomenon known
as turbophoresis. This observation was first reported by \cite{caporaloni1975transfer} in non-isotropic turbulence and later studied by \cite{reeks1983transport} in inhomogeneous turbulence.
McLaughlin (1989) observed a similar behaviour of particles in vertical-channel turbulent
flow. The author observed the near-wall accumulation of particles even in the absence of
the Saffman lift force\citep{saffman1965lift}. 
\cite{marchioli2002mechanisms} examined the effect
of vertical channel flow turbulence on particle transfer to the wall. They found that strong
coherent sweep and ejection events were the main mechanisms transferring the particles
towards the wall. \cite{sikovsky2014singularity} and \cite{johnson2020turbophoresis} pointed out the
power-law shape of particle concentrations in the viscous sublayer.

Not only do the particles migrate towards the wall, but they also show a tendency to
preferentially cluster in turbulence streaks. \cite{sardina2012wall} numerically showed that
turbophoresis and preferential clustering of particles could occur simultaneously in wall-bounded flows. \cite{squires1991preferential} numerically observed that in isotropic turbulence,
smaller particles showed a higher tendency to preferentially cluster in regions of low
vorticity and high strain rate. However, the authors assumed no effect of particles on
the flow turbulence (one-way coupling). In the review by \cite{squires1994effect}, the
authors discussed a similar effect on particles in isotropic turbulence. In wall-bounded
flow turbulence, they discussed particle clustering in low-speed streaks (LSS). A similar
observation was made by \citet{marchioli2002mechanisms} in their numerical study and by \citet{berk2023dynamics} in their experimental study of channel flows, also assuming one-way coupling between the two phases. \cite{fong2019velocity}, in their experimental
investigation of downward channel flows, also made a similar observation. \cite{ninto1996experiments} experimentally studied open-channel liquid–solid sedimenting flows. They pointed
out that particles smaller than the viscous sublayer exhibit accumulation in LSS. However,
they did not find the accumulation of larger particles along the streaks. \citet{suzuki2000simultaneous} experimentally studied the downward channel flow of water laden
with solid particles greater than twice the Kolmogorov length scale. They also noticed
the accumulation of particles in LSS. However, \citet{zhu2020interface}, in their interface-resolved simulations of upward channel flows, reported the accumulation of finite-sized
spheroidal particles in high-speed streaks (HSS). \cite{peng2024preferential}, through
their particle-resolved simulations of liquid channel flows seeded with moderate-density-ratio particles, explained that larger particles accumulate in HSS but sedimentation effects
can cause particle accumulation in LSS. They further discussed that for similar St+,
particle accumulation can be affected differently by particle–fluid density ratio and particle
diameter. Recently, \cite{dave2023mechanisms} performed simulations of horizontal channel
flows laden with high-density-ratio particles. They did not include the sedimentation effect
in their numerical study. Although the authors demonstrated the preferential clustering of
particles in LSS, the effects of particle inertia (represented by Stokes number) and mass
loading were not clear. Thus, it can be understood that the streak preference of particles for
clustering depends on the Stokes number. Nevertheless, it is still required to further this
knowledge on the combined effect of particle mass loading and Stokes number on particle
preferential clustering in a turbulent channel flow. 

The non-uniform presence of particles along the channel height can affect fluid–particle
dynamics differently at different locations. \cite{zhao2013interphasial} showed
the non-uniform particle drag on the fluid along the channel height. They argued that the
non-uniformity in the drag is due to turbophoresis. \cite{lee2015modification} observed similar
inhomogeneity in the particle drag profile. They demonstrated that higher-inertia particles
impose higher drag on the fluid near the wall. However, the authors studied the transient
effects during particle migration to the walls. Particle accumulation rates towards the wall
may differ for particles with different inertia. Thus, the transient results may not coincide
with the results of the statistically steady state. \cite{dave2023mechanisms} demonstrated
that different particle mass loadings can cause the difference in the wall accumulation of
particles even when the Stokes number is the same. Thus, it can be expected that the interphase drag profile of a particle-laden flow having the same Stokes number can be modified
by changing the particle mass loading. This is explored in the present study that further
demonstrates how this interphase drag can modulate the flow Reynolds shear stress (RSS). 

The addition of particles has been known to modulate flow turbulence. \cite{squires1994effect} showed that in isotropic turbulence, particles increased the dissipation rate, which
was higher when preferential clustering was strong. They used direct numerical simulation
(DNS) data to evaluate the two-equation $k-\epsilon$ models, and highlighted that the dissipation
was more for lighter particles that tend to exhibit higher preferential clustering. \cite{rashidi1990particle} through their experiments highlighted that larger particles
enhanced turbulence and smaller particles attenuated it in channel flows. \cite{zhao2013interphasial}, \cite{zhou2020non}, and \cite{dave2023mechanisms}, through their DNS, signified the
role of the preferential clustering of particles near the wall in modulating flow turbulence.
\cite{kulick1994particle} experimentally pointed out that the transverse fluctuations,
being at a higher frequency than the streamwise fluctuations, were attenuated by particles
because they were less responsive to high-frequency fluctuations. \cite{vreman2009two} (large-eddy simulations) and \cite{zhao2010turbulence} (DNS) observed that streamwise turbulence fluctuations were enhanced while the wall-normal and spanwise
fluctuations were damped by the particles. Although in the computational study by
\cite{zhou2020non} transverse fluctuations were damped, streamwise turbulence showed
non-monotonic behaviour along the normal distance from the wall. While there was a
suppression of streamwise fluctuations near the wall, the particles enhanced it for y+ = 40.
\cite{suzuki2000simultaneous} in their experimental study observed an increase in the fluctuations
in all directions. The higher turbulent Reynolds number (= 208) of their particle-laden
flows than that (= 172) of the clean flow could be the reason for the enhanced fluctuations
even in the transverse directions. In contrast, \cite{kussin2002experimental} found that the
streamwise turbulence was suppressed more than the wall-normal turbulence. However,
their experimental study included various effects like gravity in a horizontal channel,
which can cause a loss in particle concentration symmetry across the channel width and
affect particle motion. Moreover, the deviation in the particle diameter from the mean
diameter was at least $\pm 30 \mu m$. All these factors may complicate the analysis, and thus
a conclusive interpretation of the influence that dispersed particles have on modulating
turbulent fluctuations remains elusive.

In the present study, the use of the quadrature-based moment method \citep{desjardins2006quadrature,desjardins2008quadrature} has been extended to simulate dispersed particle-laden turbulent
flows in a horizontal channel using the finite-volume method. By coupling the particle
solver with a compressible solver for the fluid phase, we aim to demonstrate the
effectiveness of this Eulerian method of moments in simulating dispersed particle fields in
various configurations and environments in both low-speed and high-speed compressible
flows. Our results show, for the first time, that this framework accurately captures the
intrinsic behaviour of particles and their influence on fluid dynamics, focusing on key
phenomena related to fluid–particle interactions, such as particle migration towards
the wall, preferential clustering of particles within turbulent streaks, fluid turbulence
modulation, particle mass flow rate and the drag exerted by particles on the fluid.
Furthermore, we investigate how varying the Stokes number and particle mass loading
impacts these phenomena without requiring any statistical models for particle properties,
as is typically needed in Lagrangian approaches. We demonstrate that the Eulerian
quadrature-based moment method effectively captures the effects of Stokes number and
particle mass loading on these interactions.

The paper is organised as follows. The governing equations are described in \textsection~\ref{sec:num_proc}, which
outlines the fluid transport equations and the particle moment transport equations. The
governing equations are solved using finite-volume schemes for the two phases, which are
described in \textsection~\ref{sec:numerical_method}. Section \ref{sec:flow_config} describes the flow configuration, illustrating various flow parameters of the two phases. In \textsection~\ref{subsec:flow_stats}, we show various turbulence statistics of particle-free
channel (PFC) and various particle-laden channel (PLC) flows. The schemes and methodologies for the particle-free and PLC flows are also validated in this section. Section \ref{subsec:parclus}
identifies the particle migration towards the wall and the preferential clustering of particles
in LSS. Section \ref{subsec:accu_eff} discusses the effects of particle accumulation and their preferential
clustering on various flow phenomena like particle mass flow rate (\textsection~\ref{subsubsec:parmass}), flow RSS and
interphase drag (\textsection~\ref{subsubsec:rss}) and fluid turbulence modulation through the interaction between
particle streaks and fluid velocity streaks (\textsection~\ref{subsubsec:tur_mod}). Conclusions are drawn in \textsection~\ref{sec:concl}.

\section{\label{sec:num_proc}Governing equations}

\quad The fluid is modelled as an ideal compressible gas, whereas the particle phase is treated as a solid and dilute phase with no inter-particle interactions. The particle volume fraction is thus neglected in the convective terms of the carrier flow. The Stokes number is assumed to be small, allowing for the formulation of different particle moments \citep{desjardins2008quadrature}. Although the particles are in dilute concentrations, significant mass loadings of particles ensure the presence of two-way coupling between the two phases. The particle Reynolds number $Re_p$ is assumed to be in the Stokes regime. Thus, the two phases are coupled using the Stokes drag. Gravity is neglected, and elastic collisions are assumed between the particles and the wall. The particles are assumed to be in thermal equilibrium with the fluid at all times, so heat transfer between the two phases is not resolved. Finally, the flow Prandtl number is set to 0.71.

\subsection{\label{subsec:fluid_eq}Fluid phase transport equations}
\quad The following compressible transport equations for the carrier fluid phase are solved:
\begin{align}
	\frac{\partial \rho}{\partial t} + \vec{\nabla}.(\rho\vec{V}) &= 0
	\nonumber\\
	\frac{\partial (\rho u)}{\partial t} + \vec{\nabla}.(\rho u \vec{V} + P \hat{x}) &= \vec{\nabla}.(\tau_{xx}\hat{x} + \tau_{xy}\hat{y} + \tau_{xz}\hat{z}) - F_{Dx}
	\nonumber\\
	\frac{\partial (\rho v)}{\partial t} + \vec{\nabla}.(\rho v \vec{V} + P \hat{y}) &= \vec{\nabla}.(\tau_{xy}\hat{x} + \tau_{yy}\hat{y} + \tau_{yz}\hat{z}) - F_{Dy} \label{eq:fluid_trans}\\
	\frac{\partial (\rho w)}{\partial t} + \vec{\nabla}.(\rho w \vec{V} + P \hat{z}) &= \vec{\nabla}.(\tau_{xz}\hat{x} + \tau_{yz}\hat{y} + \tau_{zz}\hat{z}) - F_{Dz}
	\nonumber\\
	\frac{\partial (\rho e)}{\partial t} + \vec{\nabla}.((\rho e + P) \vec{V}) &= \vec{\nabla}.(q_x\hat{x} +  q_y\hat{y} + q_z\hat{z}) - E_D \nonumber
\end{align}

\noindent  $\vec{F}_D (= F_{Dx}\hat{x} + F_{Dy}\hat{y} + F_{Dz}\hat{z})$ represents the interphase drag force per unit volume. Since the particle transport system considers the particle parameter values at two nodes within the control volume (explained in \textsection \ref{subsec:part_eq}), the equivalent values of these parameters, $\rho_p$ and $\vec{U}_p$, should be considered for the drag calculation in the fluid phase. Assuming a very low particle $Re_p$, Stokes drag has been considered here, $\vec{F}_{D} = 3\pi \mu d_{p}(\vec{U} - \vec{U}_p)\frac{\rho_{p}}{m_p}$; $m_p$ and $d_p$, respectively represent the mass and diameter of a single particle. $E_D = \vec{F}_{D}.\vec{U}_p$, is the energy transfer rate due to the drag force between the two phases. $\tau$ represents the viscous stress tensor and $q$ is the energy flux rate due to fluid viscous forces and heat diffusion.

\subsection{\label{subsec:part_eq}Particle moment transport equations}

\quad The conventional transport method for Eulerian particles averages out particle velocities, which can nullify the reflective effect of the wall. This method may also average out the distinct velocities of particles within the control volume, leading to inaccuracies in cases involving inter-particle collisions, trajectory crossing and dispersion. Furthermore, the absence of a pressure term in the transport equations can cause ill-posedness in the system \citep{bouchut1994zero}. This pressureless gas dynamics approach can result in an overestimation of the particle number density in LSS of turbulent flows \citep{desjardins2008quadrature}. 

To improve the accuracy of the simulations, a two-node quadrature-based moment method, as proposed by \cite{desjardins2006quadrature,desjardins2008quadrature}, is employed in the present study. The use of two nodes prevents complete averaging of the velocity field and retains the reflected velocity of particles at the wall, thereby accurately capturing particle-wall interactions. This method has also been shown to accurately capture particle clustering in turbulent flows. Additionally, the study considers transport equations for moments up to third order, enhancing the method’s ability to effectively capture particle trajectory crossing within the control volume \citep{fox2012large,baker2020verification}. Since the quadrature method accounts for velocities at two distinct nodes within the control volume, it is also a promising approach for simulating relatively higher-Mach-number flows while avoiding numerical instabilities and negative weights \citep{fox2012large}. However, the method comes with certain limitations; due to the field assumption for the particle phase, the method may fail to solve flows with high Stokes numbers and low particle mass loadings, which can result in very sparse population of particles in a control volume. A division by a very small value of the particle number density can ultimately lead to failure of the method. Furthermore, the method has not been validated for high-Mach-number flows with higher hyperbolic requirements.

For a non-evaporating, monodispersed particle phase with no inter-particle collisions, turbulent dispersion or atomisation, the quadrature-based moment method can be derived from the Williams equation \citep{williams1958spray}: 
\begin{equation}
    \frac{\partial f}{\partial t} + \vec{\nabla}_E.(f \vec{U}_p) + \vec{\nabla}_v.\bigg(f\frac{\vec{F}_D}{m_p}\bigg) = 0.
\end{equation}

\noindent Here, $f$ represents the number density function, $\vec{\nabla}_E$ represents the normal gradient in the Euclidean space and $\vec{\nabla}_v$ represents the gradient in the velocity space. Since the particle density is considered to be much higher than the fluid density, only the drag force is considered between the two phases. 

The quadrature-based moment method solves the system of equations that conserves up to the third-order moment of velocities within the control volume. For small Stokes numbers, these moments can be approximated using the particle parameters at two nodes: 
\begin{align}
   \text{zeroth-order moment: } M^0~~~ &= ~\rho_{p1} + \rho_{p2} \nonumber\\
   \text{first-order moment: } M^1_i~~~ &= ~\rho_{p1}U_{p1i} + \rho_{p2}U_{p2i} \nonumber\\
   \text{second-order moment: } M^2_{ij}~~ &= ~\rho_{p1}U_{p1i}U_{p1j} + \rho_{p2}U_{p2i}U_{p2j}\\
   \text{third-order moment: } M^3_{ijk} &= ~\rho_{p1}U_{p1i}U_{p1j}U_{p1k} + \rho_{p2}U_{p2i}U_{p2j}U_{p2k} \nonumber\\
   \text{third-order moment: } Q~~~~~ &= ~\rho_{p1}\sum\limits_{x,y,z}U_{p1}^3 + \rho_{p2}\sum\limits_{x,y,z}U_{p2}^3 \nonumber
\end{align}
\noindent Here, $i,j,k$ represent the three directions and $U_{p1}$ and $U_{p2}$ are the particle velocities at node 1 and node 2, respectively. Similarly, $\rho_{p1}$ and $\rho_{p2}$, different from the particle material density, represent the equivalent densities of particles at the two nodes within the control volume. Using these definitions of the moments, the moment transport equations for a three-dimensional particle flow can be formulated as
 \begin{align}
	\frac{\partial M^0}{\partial t} + \vec{\nabla}.(M^1_x\hat{x} + M^1_y\hat{y} + M^1_z\hat{z}) &= 0 \nonumber\\
	\frac{\partial M^1_x}{\partial t} + \vec{\nabla}.(M^2_{xx}\hat{x} + M^2_{xy}\hat{y}  + M^2_{xz}\hat{z}) &= \frac{\rho_{p1}}{m_p}F_{Dx1} + \frac{\rho_{p2}}{m_p}F_{Dx2} \nonumber\\
	\frac{\partial M^1_y}{\partial t} + \vec{\nabla}.(M^2_{yx}\hat{x} + M^2_{yy}\hat{y} + M^2_{yz}\hat{z}) &= \frac{\rho_{p1}}{m_p}F_{Dy1} + \frac{\rho_{p2}}{m_p}F_{Dy2} \nonumber\\
	\frac{\partial M^1_z}{\partial t} + \vec{\nabla}.(M^2_{zx}\hat{x} + M^2_{zy}\hat{y} + M^2_{zz}\hat{z}) &= \frac{\rho_{p1}}{m_p}F_{Dz1} + \frac{\rho_{p2}}{m_p}F_{Dz2} \nonumber\\
	\frac{\partial M^2_{xx}}{\partial t} + \vec{\nabla}.(M^3_{xxx}\hat{x} + M^3_{xxy}\hat{y} + M^3_{xxz}\hat{z}) &= 2\frac{\rho_{p1}}{m_p}u_{p1}F_{Dx1} + 2\frac{\rho_{p2}}{m_p}u_{p2}F_{Dx2}\label{eq:part_trans}\\
    \frac{\partial M^2_{yy}}{\partial t} + \vec{\nabla}.(M^3_{yyx}\hat{x} + M^3_{yyy}\hat{y} + M^3_{yyz}\hat{z}) &= 2\frac{\rho_{p1}}{m_p}v_{p1}F_{Dy1} + 2\frac{\rho_{p2}}{m_p}v_{p2}F_{Dy2} \nonumber\\
    \frac{\partial M^2_{zz}}{\partial t} + \vec{\nabla}.(M^3_{zzx}\hat{x} + M^3_{zzy}\hat{y} + M^3_{zzz}\hat{z}) &= 2\frac{\rho_{p1}}{m_p}w_{p1}F_{Dz1} + 2\frac{\rho_{p2}}{m_p}w_{p2}F_{Dz2} \nonumber
 \end{align}
 \begin{align}
	\frac{\partial Q}{\partial t} + \vec{\nabla}.(R_x\hat{x} + R_y\hat{y} + R_z\hat{z}) &= 3\frac{\rho_{p1}}{m_p}\sum\limits_{x,y,z}(U_{p1}^2F_{D1}) + 3\frac{\rho_{p2}}{m_p}\sum\limits_{x,y,z}(U_{p2}^2F_{D2}) \nonumber
\end{align}
$\vec{F}_{D1} = F_{Dx1}\hat{x} + F_{Dy1}\hat{y} + F_{Dz1}\hat{z}$ and $\vec{F}_{D2} = F_{Dx2}\hat{x} + F_{Dy2}\hat{y} + F_{Dz2}\hat{z}$ represent the drag forces between the fluid and the particle field, at the two nodes. At the $n$th node,
$\vec{F}_{Dn} = 3\pi \mu d_{p}(\vec{U} - \vec{U}_{pn})$. The last equation in the system represents the transport equation for the third-order moment $Q$ with $\vec{R} = R_x\hat{x} + R_y\hat{y} + R_z\hat{z}$ being the closure required for the system. In the $i$th direction, the value of $R$ reads $R_i = \rho_{p1}\bigg(\sum\limits_{x,y,z}U_{p1}^3\bigg)U_{p1i} + \rho_{p2}\bigg(\sum\limits_{x,y,z}U_{p2}^3\bigg)U_{p2i}$.

By solving the above system, the values of various moments within the control volume can be evaluated. Using these values, equivalent values of the particle velocity and equivalent density inside the control volume are calculated as $\rho_p = M^0$, $u_p = M^1_x/M^0$, $v_p = M^1_y/M^0$, and $w_p = M^1_z/M^0$.

\section{\label{sec:numerical_method}Numerical methods}

\quad The set of equations in the system (\ref{eq:fluid_trans}) and (\ref{eq:part_trans}) can be summarised as
\begin{equation}
    \frac{\partial C}{\partial t} + \vec{\nabla}.\vec{F} = \vec{\nabla}.\bar{\bar{S}}_v + S_c~.
\end{equation}
For a control volume $\Delta\mathcal{V}$, the system can be written as
\begin{equation}
    \frac{\partial C}{\partial t}\Delta \mathcal{V} + \vec{F}.d\vec{s} = \bar{\bar{S}}_v.d\vec{s} + S_c\Delta \mathcal{V}~,
    \label{eq:transport_model}
\end{equation}
where $C$ represents the conservative variables, $\vec{F}.\vec{ds}$ is the flux of convective variables, $\bar{\bar{S}}_v.d\vec{s}$ is the viscous source term in the fluid transport equation, and $S_c\Delta \mathcal{V}$ represents the interphase coupling source term. The matrices representing these variables are detailed in the appendix\ref{app:matrix}.

The numerical method is similar to that described by \cite{desjardins2008quadrature}, where they considered a fractional two-step approach to solve the particle moment equations. The approach has been extended to solve the fluid transport equations in the present work. In the first step, the transport equations of both phases described by the system \ref{eq:transport_model} are solved without the coupling source terms $S_c\Delta \mathcal{V}$ to obtain the approximate values of the conservative variables $C^*$ after the time step $\Delta t$. A Runge-Kutta type integration method is used to calculate $C^*$. In the second step, the system \ref{eq:transport_model} is solved without $\vec{F}.\vec{ds}$ and $\bar{\bar{S}}_v.d\vec{s}$, using the values approximated in the first step, to obtain new values of the conservative variables $C^{**}$. 

To increase the temporal accuracy of the solution to second order, the mid-time-step values are estimated by calculating the mean values of $C^{**}$, calculated after the time step $\Delta t$, and $C$, from the previous time step. Using these mid-time-step values and the previous time values, the fractional two-step approach is repeated to calculate the final values of the conservative variables that are second-order accurate in time.

The flux of the fluid convective variables $\vec{F}_f.d\vec{s}$ is calculated explicitly using the simple low dissipation (advection upstream splitting method (AUSM)) scheme (SLAU2) \citep{kitamura2010improvements,kitamura2013towards}. The scheme belongs to the AUSM family of schemes that were originally developed to simulate high-speed flows and have since been modified over the years to simulate both high and low-speed flows \citep{liou1993new,liou1996sequel,liou2006sequel,shima2011parameter,shima2013improvement,chen2020improved}. For the present simulations, the scheme is converted to second order in space by interpolating the fluid parameters at the cell face using the least squares method \citep{bjorck1996numerical} and the Venkatakrishnan limiter \citep{venkatakrishnan1993accuracy,venkatakrishnan1995convergence}. The same least squares method is also used to evaluate the stress tensors $\bar{\bar{S}}_v$ for calculating the viscous source terms $\bar{\bar{S}}_v.d\vec{s}$. As an upwind scheme, SLAU2 can be overly dissipative for low-Mach-number flows where the dissipation rate increases as the inverse of the Mach number, $M$ \citep{thornber2008entropy}. Thus, in the incompressible regime, for $M<0.3$, a significant dissipation of the solution can be expected. To deal with this inherent dissipation of the scheme, a velocity correction proposed by \cite{thornber2008improved} is implemented. The correction method was applied by \cite{kokkinakis2009investigation} in the simulation of a turbulent channel flow at $M = 0.2$ using the upwind Harten–van Leer–Lax contact scheme \citep{toro1994restoration,toro2019hllc}. The correction significantly reduced the dissipation on coarser grids. \cite{matsuyama2014performance} demonstrated the effectiveness of the correction in reducing the dissipation in his DNS simulation of a low-Mach-number channel flow using various upwind schemes, including SLAU (predecessor of SLAU2) \citep{shima2011parameter}. According to the correction, the velocity values extrapolated to the left and right of the cell face ($U_l$ and $U_r$) are adjusted using the minimum value of local interface Mach numbers ($M_l$ and $M_r$):
\begin{equation}
    U_l^{LM} = \frac{U_l + U_r}{2} + Z\frac{U_l - U_r}{2};\quad U_r^{LM} = \frac{U_l + U_r}{2} - Z\frac{U_l - U_r}{2}
\end{equation}
\begin{equation}  Z = min[1, max(M_l, M_r)] \end{equation}

\noindent This adjustment of the velocities is incorporated after the calculation of the mass flux and the pressure flux at the cell face. The adjustment reduces sharp changes in the left and right values of the velocities, which contribute to the dissipation. We find that using the absolute values of the local Mach numbers to calculate $Z$ gives the best results. As shown in \textsection \ref{subsec:flow_stats}, the correction significantly improved the results of the low-Mach-number flow when the fluid is in an incompressible regime. For high-Mach-number flows, $Z$ takes the value of unity, thus retaining the original scheme.

The flux values of the particle convective variables $\vec{F}_p.d\vec{s}$ are calculated using an upwind-type-flux splitting method described by \cite{desjardins2006quadrature,desjardins2008quadrature}. The method is based on the kinetic schemes \citep{pullin1980direct,deshpande1986second,perthame1992second,estivalezes1996high}. At each cell face, the scheme decides the contribution of the parent ($l$) and the neighbour ($r$) cells based on the sign of the particle mass flow rate ($\dot{m}_p = \rho_p \vec{U}_p.d\vec{s}$) at each node. For the $n$th node, we have
\begin{align}
    \dot{m}_{pln} = \rho_{pln} \times max(\vec{U}_{pln}.d\vec{s},0);\quad 
    \dot{m}_{prn} = \rho_{prn} \times min(\vec{U}_{prn}.d\vec{s},0)
\end{align}

Using the particle mass flux, the convective flux values at the cell face are given by 
\begin{align}
    [\vec{F}.d\vec{s}]_l = \dot{m}_{pl1}\begin{bmatrix}
        1\\
        u_{pl1}\\
        v_{pl1}\\
        w_{pl1}\\
        u_{pl1}\times u_{pl1}\\
        v_{pl1}\times v_{pl1}\\
        w_{pl1}\times w_{pl1}\\
        \sum\limits_{x,y,z}U_{pl1}^3\\
    \end{bmatrix} + \dot{m}_{pl2}\begin{bmatrix}
        1\\
        u_{pl2}\\
        v_{pl2}\\
        w_{pl2}\\
        u_{pl2}\times u_{pl2}\\
        v_{pl2}\times v_{pl2}\\
        w_{pl2}\times w_{pl2}\\
        \sum\limits_{x,y,z}U_{pl2}^3\\
    \end{bmatrix}
\end{align}
\begin{align}
    [\vec{F}.d\vec{s}]_r =  \dot{m}_{pr1}\begin{bmatrix}
        1\\
        u_{pr1}\\
        v_{pr1}\\
        w_{pr1}\\
        u_{pr1}\times u_{pr2}\\
        v_{pr1}\times v_{pr2}\\
        w_{pr1}\times w_{pr2}\\
        \sum\limits_{x,y,z}U_{pr2}^3\\
    \end{bmatrix}+ \dot{m}_{pr2}\begin{bmatrix}
        1\\
        u_{pr2}\\
        v_{pr2}\\
        w_{pr2}\\
        u_{pr2}\times u_{pr2}\\
        v_{pr2}\times v_{pr2}\\
        w_{pr2}\times w_{pr2}\\
        \sum\limits_{x,y,z}U_{pr2}^3\\
    \end{bmatrix}
\end{align}

\begin{align}
    \vec{F}.d\vec{s} = [\vec{F}.d\vec{s}]_l + [\vec{F}.d\vec{s}]_r
\end{align}

The particle variables at each node can be given by
\begin{align}
    \rho_{p1} = (1/2 + x)M^0;\quad \rho_{p2} = (1/2 - x)M^0
\end{align}
\begin{align}
    U_{p1i} = U_{pi} - \sqrt{(\rho_{p2}/\rho_{p1})}\sigma_{pi};\quad U_{p2i} = U_{pi} + \sqrt{(\rho_{p1}/\rho_{p2})}\sigma_{pi}
\end{align}

Here,
\begin{align}
    x=\frac{q_p/2}{\sqrt{q_p^2+4\bigg(\sum\limits_{i=x,y,z}\sigma_{pi}\bigg)^2}};\quad U_{pi} = M_i^1/M^0;\quad \sigma_{pi} = \sqrt{\frac{M^0M_{ii}^2 - (M_{i}^1)^2}{(M^0)^2}}
\end{align}
\begin{align}
    q_p = \frac{1}{M_0}\bigg(Q - M^0\sum\limits_{i=x,y,z}U_{pi}^3 - 3M^0\sum\limits_{i==x,y,z}\sigma_{pi}^2U_{pi}\bigg)
\end{align}

The scheme is converted to second order in space by interpolating all convective variables at the cell boundary in a fashion similar to that done for the fluid. Using the values of particle variables at the two nodes, the calculation of coupling source terms within a cell is straightforward from (\ref{eq:part_trans}). This stable numerical scheme for the particle phase should account for the hyperbolicity of the particle transport system \citep{desjardins2008quadrature}.

A GPU-accelerated in-house numerical solver is developed for DNS of turbulent flows laden with dispersed particles. The whole methodology is developed in CUDA \citep{sanders2010cuda,cook2012cuda} to implement GPU parallelisation, which significantly reduces computation time. The computations are carried out using multiple A100 graphical processor units in a DGX cluster. 

\section{\label{sec:flow_config} Flow parameters and configuration}

\quad The particle volume fraction $\phi_v$, defined as the total volume of particles per unit volume of fluid, and the particle mass loading $\phi_m$, defined as the ratio between the total particle mass in the channel and the total fluid mass, determine the type of interactions between the two phases \citep{Balachandar2010,kasbaoui2019turbulence}. At the limit of low values of $\phi_v$ and $\phi_m$, the particles act as passive tracers, hardly affecting the flow or other particles. At $\phi_m =$ $O(1)$ and higher values of $\phi_v$ ($> 10^{-3}$), both particle-fluid and particle-particle interactions are significant, but when $10^{-6} < \phi_v < 10^{-3}$, particle-particle interactions become secondary or negligible and particle-fluid interactions dominate. The nature of particle-fluid interactions is determined by the Stokes number $St$. Since two-way coupling is considered in the present work, the flow configuration parameters for the two phases are chosen in such a way that they ensure the presence of fluid-particle interactions but justify the omission of the particle-particle interactions.

\begin{table}
		\centering
		\begin{tabular}{p{0.07\linewidth} p{0.07\linewidth} p{0.06\linewidth} p{0.06\linewidth} p{0.1\linewidth} p{0.1\linewidth} p{0.1\linewidth} p{0.13\linewidth}}
		$St^+$ & $\phi_m$  & $l_x/h$ & $l_y/h$ & $l_z/h$ & $d_p/h$ & $\rho_{pp}/\rho$ & $\phi_{v0} (\times 10^{-5})$ \\
			& & & & & & & \\
            6 & 0.2 & $4\pi$ & $2$ & $(4/3)\pi$ & 0.0005 & 13697 & 1.46 \\
            6 & 0.6 & $4\pi$ & $2$ & $(4/3)\pi$ & 0.0005 & 13697 & 4.38 \\
            6 & 1.0 & $4\pi$ & $2$ & $(4/3)\pi$ & 0.0005 & 13697 & 7.08 \\
            30  & 0.2 & $12$ & $2$ & $6$ & 0.0011 & 13697 & 1.45 \\
            30  & 0.6 & $12$ & $2$ & $6$ & 0.0011 & 13697 & 4.37 \\
		  30  & 1.0 & $12$ & $2$ & $6$ & 0.0011 & 13697 &  7.05 \\ 
	\end{tabular}
    \caption{Table representing the non-dimensional parameters for particle-laden turbulent channel flows. Here, $\rho_{pp}$ and $\rho$ are the density of the particle material and the fluid, respectively, and $\phi_{v0}$ is the initial volume fraction of particles when they are introduced uniformly in the channel. The channel dimensions in the streamwise, wall-normal, and spanwise directions are $l_x$, $l_y$, and $l_z$, respectively.}
	\label{tab:param}
	\end{table}
 
    Monodisperse PLC flows are considered in the study. Two different frictional Stokes numbers $St^+$ are studied, where $St^+ = \tau_p u_{\tau}^2/\nu$ is based on the frictional time scale of the flow. Here, $\tau_p$ is the particle time scale, $u_\tau$ is the frictional velocity ($u_\tau = Re_\tau\times\nu/h$), and $\nu$ represents the kinematic viscosity of the fluid. Three different values of $\phi_m$ are considered for each particle type. These, along with other parameters, are outlined in table \ref{tab:param}. In all cases, the particle size is less than the Kolmogorov length scale of the carrier flow. The number density of particles inside the domain is chosen such that $\phi_{v0}$ is between $10^{-5}$ and $10^{-4}$, thus ensuring the dominance of the two-way coupling. We also find that even in clustered regions, $\phi_v$ is always $< 10^{-3}$. Hence, we expect that in such clustered regions particle interactions may still be neglected.

For each case, the fluid flow inside the channel has $Re = 2800$ and $Re_\tau = 180$, based on the channel half height. The fluid velocity corresponds to $M = 0.12$. Thus, the flow can be considered incompressible. By simulating an incompressible fluid flow using a compressible flow solver, we can thus establish the capability of the developed solver to effectively solve the flow in a wide range of Mach numbers. However, it should be mentioned that the solver assumes the fluid to be an ideal gas. 

\begin{figure}
		\begin{subfigure}{0.49\textwidth}
			\begin{minipage}[]{0.05\linewidth}
                    \subcaption{}
                \end{minipage}

                \begin{minipage}[]{1\linewidth}
                    \includegraphics[clip=true, trim = 0.0in 0.0in 0.0in 0.0in,width=0.9\textwidth]{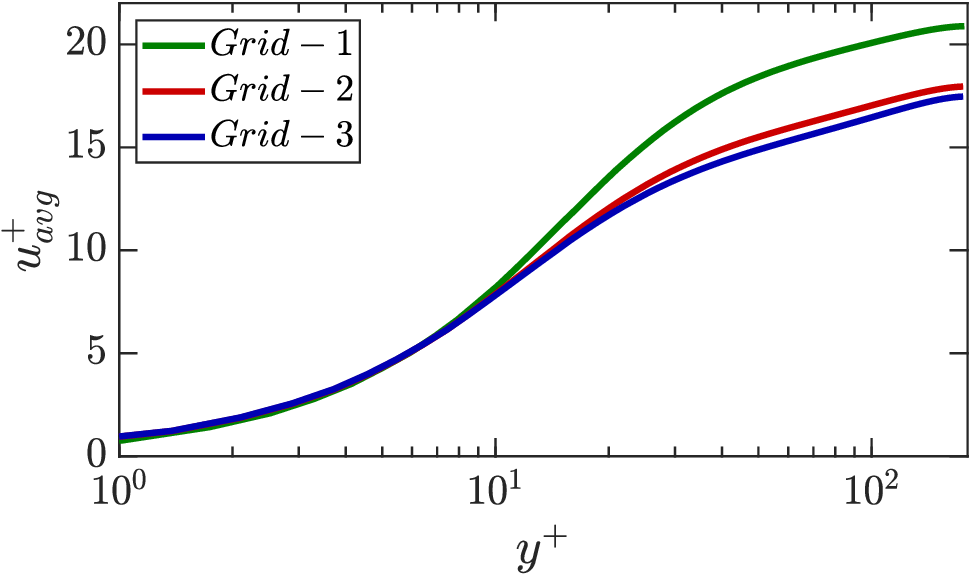}
                \end{minipage}
			\label{fig:gis_avg_u}
		\end{subfigure}
		\begin{subfigure}{0.49\textwidth}
			\begin{minipage}[]{0.05\linewidth}
                    \subcaption{}
                \end{minipage}

                \begin{minipage}[]{1\linewidth}
                    \includegraphics[clip=true, trim = 0.0in 0.0in 0.0in 0.0in,width=0.9\textwidth]{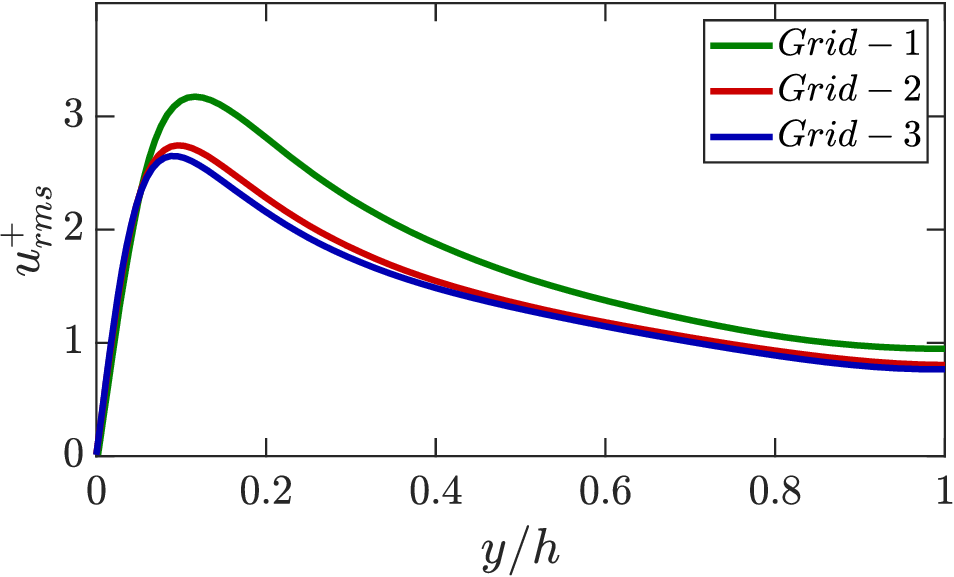}
                \end{minipage}
			\label{fig:gis_rms_u}
		\end{subfigure}

             \begin{subfigure}{0.49\textwidth}
			\begin{minipage}[]{0.05\linewidth}
                    \subcaption{}
                \end{minipage}

                \begin{minipage}[]{1\linewidth}
                    \includegraphics[clip=true, trim = 0.0in 0.0in 0.0in 0.0in,width=0.9\textwidth]{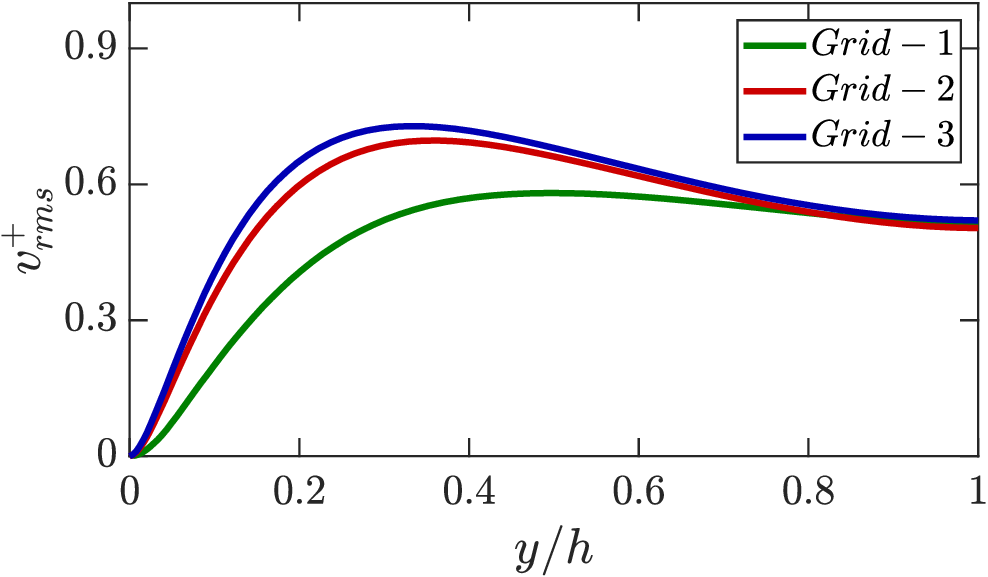}
                \end{minipage}
			\label{fig:gis_rms_v}
		\end{subfigure}
		\begin{subfigure}{0.49\textwidth}
			\begin{minipage}[]{0.05\linewidth}
                    \subcaption{}
                \end{minipage}

                \begin{minipage}[]{1\linewidth}
                    \includegraphics[clip=true, trim = 0.0in 0.0in 0.0in 0.0in,width=0.9\textwidth]{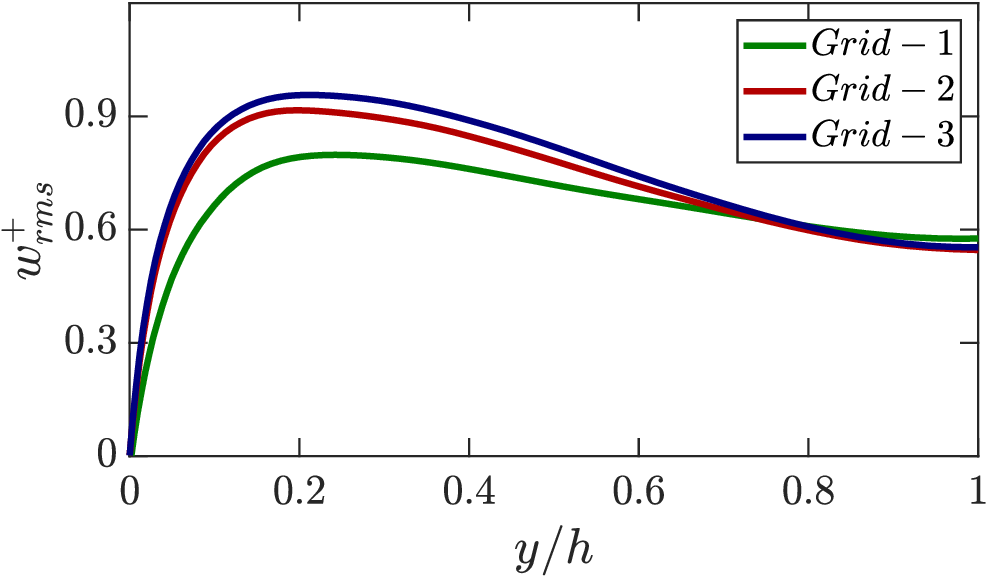}
                \end{minipage}
			\label{fig:gis_rms_w}
		\end{subfigure}

             \caption{Variation of (a) mean streamwise velocity, and rms of velocity fluctuations in (b) streamwise, (c) wall-normal,  and (d) spanwise directions along the channel height in a particle-free turbulent channel flow. Results for the three grid types are compared for grid sensitivity study. Here, $\Delta x_{grid-1} = 2\Delta x_{grid-2} = 4\Delta x_{grid-3}$; $\Delta z_{grid-1} = 2\Delta z_{grid-2} = 4\Delta z_{grid-3}$.}
		\label{fig:gis}
	\end{figure}
    A rectangular channel is used for the simulations. A slightly wider domain is considered for the high $St^+$ case to capture the increased spanwise spacing of the particle clusters \citep{dave2023mechanisms}. Although the domain of high $St^+$ is wider than that of low $St^+$, comparing the effects of two particle types is plausible due to the use of periodic boundary conditions in the spanwise direction. The smaller domain for the low $St^+$ case is similar to the domain used by \cite{moser1999direct} in their DNS of incompressible fluid flow. \cite{dave2023mechanisms} used a similar domain in their DNS of PLC flows with $St^+ = 6$ using the Lagrangian approach. The larger domain is similar to that used by \cite{zhao2010turbulence}, \cite{zhao2013interphasial} and \cite{zhou2020non} in their Lagrangian DNS of PLC flows with $St^+ = 30$. Similar domains allow us to effectively compare and validate the results of our simulation employing a compressible fluid solver and the Eulerian particle method. For both domains, the mesh resolution is constant in the streamwise and spanwise directions. In the wall-normal direction, the mesh is refined smoothly from the channel centre, where $\Delta y = 0.028h$, towards the wall, where $\Delta y = 0.0036h$ resulting in a $y^+ = 0.65$. While the mesh resolution in the wall-normal direction is based on previous studies \citep{zhao2013interphasial,dave2023mechanisms} and is highly refined, the resolution in the streamwise and spanwise directions was adopted after conducting a grid sensitivity analysis involving three grid types such that $\Delta x$ and $\Delta z$ of the successive grid type are half that of the previous grid type. Figure \ref{fig:gis} demonstrates the flow statistics in the PFC flow using the three grid types. Here, grid-1 is the most coarse, while grid-3 is the most refined. From the figure, it can be concluded that the results show grid convergence for grid-2 and grid-3. Using grid-3 does not give any significant change in the results for the increased number of elements. Thus, we have considered grid-2 for the present study. This grid type has $\Delta x = 0.048h$ and $\Delta z = 0.032h$.
    
    Periodic boundary conditions are applied in the streamwise and spanwise directions for both phases. A no-slip condition is imposed on the fluid at the wall. The elastic reflection condition for particles at the wall is defined by balancing the inflow and outflow of particle flux, ensuring no net exchange of mass or tangential particle momentum through the wall while reversing the normal component of the particle momentum. This type of boundary condition has been utilised by \cite{desjardins2006quadrature,desjardins2008quadrature} to demonstrate the effectiveness of the quadrature-based moment method in capturing wall reflections of the Eulerian particle phase. A small pressure difference between the inlet and the outlet is introduced to balance the wall shear stress caused by viscosity, thus maintaining the flows at $Re_\tau = 180$.

The domains are initialised with the fluid-phase variables without the particles. A zero mean velocity of the flow is considered in the spanwise and wall-normal directions. The velocity field in the streamwise direction is initialised in such a manner that the flow in the near-wall region ($|y| < 0.9h$) is in the negative streamwise direction, while the flow in the remaining region is in the positive streamwise direction. This initialisation forces strong shear flows in the near-wall regions, which generate strong disturbances that serve as initial perturbations in the flow. Although random fluctuations are also introduced in the three velocity components, their effect on the development of turbulence was found to be insignificant. A similar observation was also made by \cite{matsuyama2014performance}, who used the same technique to initialise the turbulent channel flow.

The PFC flow in each case is then simulated for around 140 eddy turnover times $\Delta t_\eta$ $(= \frac{h}{u_\tau}$) for the flow to reach a statistically steady state. The flow is further simulated for 60 $\Delta t_\eta$ to record the data for analysis. Subsequently, the particles are introduced uniformly into the flow with the same velocity as the fluid. The PLC flows are then simulated for around 140$\Delta t_\eta$ to achieve another statistically steady state, after which data are recorded in the same fashion as for the PFC flow.
    
\section{\label{sec:res_obs} Results and discussion}
\subsection{\label{subsec:flow_stats} Flow statistics}
\quad The PFC flow is used here and in subsequent sections as a baseline reference, which is compared with the PLC flow cases to highlight the mean flow and turbulence modulations induced by the particles. 
As the objective of this study is to expand the capabilities of the Eulerian approach, it is instructive to compare the results with those of previous studies, most of which focus on the low Mach number regime.
To achieve this, we start by assessing the ability of the SLAU2 scheme (second order) to simulate the PFC flow within the incompressible regime.

The SLAU2 scheme is an upwind scheme that can be used to simulate high-speed flows \citep{kitamura2013towards,kitamura2016reduced,mamashita2021slau2}. Although its ability to simulate low-speed flows has been claimed, it has not been validated for channel flows. \cite{matsuyama2014performance} established the validity of its predecessor (SLAU) to simulate low-speed flows. However, the author used up to seventh-order weighted essentially non-oscillatory (WENO) interpolation for the validation. Yet, achieving a seventh-order conversion demands substantial computational resources. The second-order SLAU2 scheme, on the other hand, is expected to be more efficient and computationally less expensive.

        \begin{figure}
        \begin{subfigure}{0.49\textwidth}
			\begin{minipage}[]{0.05\linewidth}
                    \subcaption{}
                \end{minipage}

                \begin{minipage}[]{1\linewidth}
                    \includegraphics[clip=true, trim = 0.0in 0.0in 0.0in 0.0in,width=0.9\textwidth]{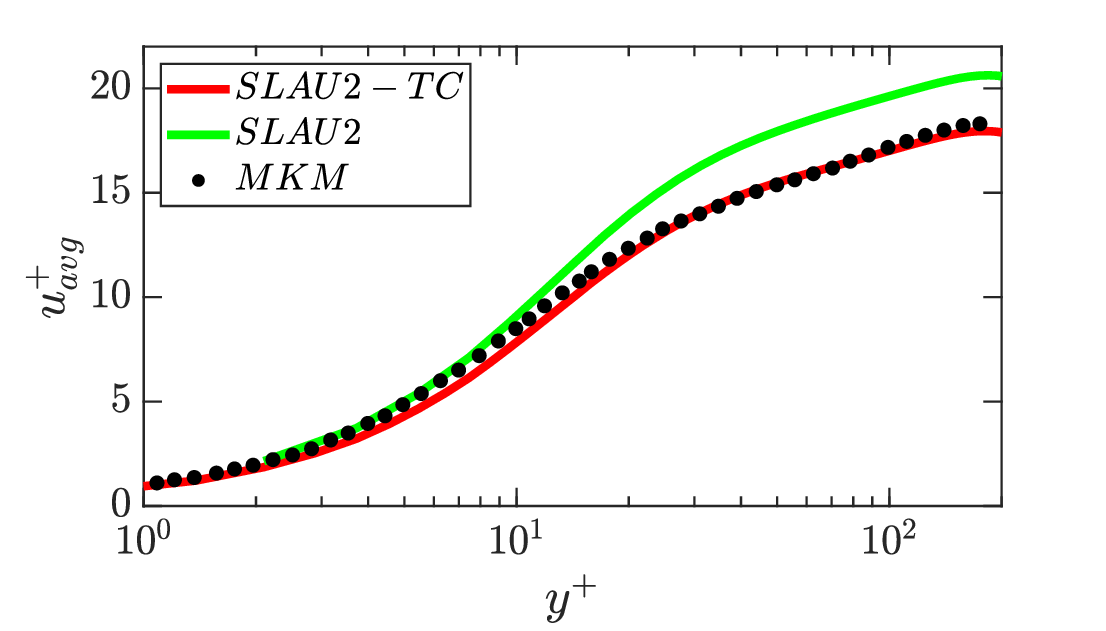}
                \end{minipage}
			\label{fig:avg_val}
		\end{subfigure}
		\begin{subfigure}{0.49\textwidth}
			\begin{minipage}[]{0.05\linewidth}
                    \subcaption{}
                \end{minipage}

                \begin{minipage}[]{1\linewidth}
                    \includegraphics[clip=true, trim = 0.0in 0.0in 0.0in 0.0in,width=0.9\textwidth]{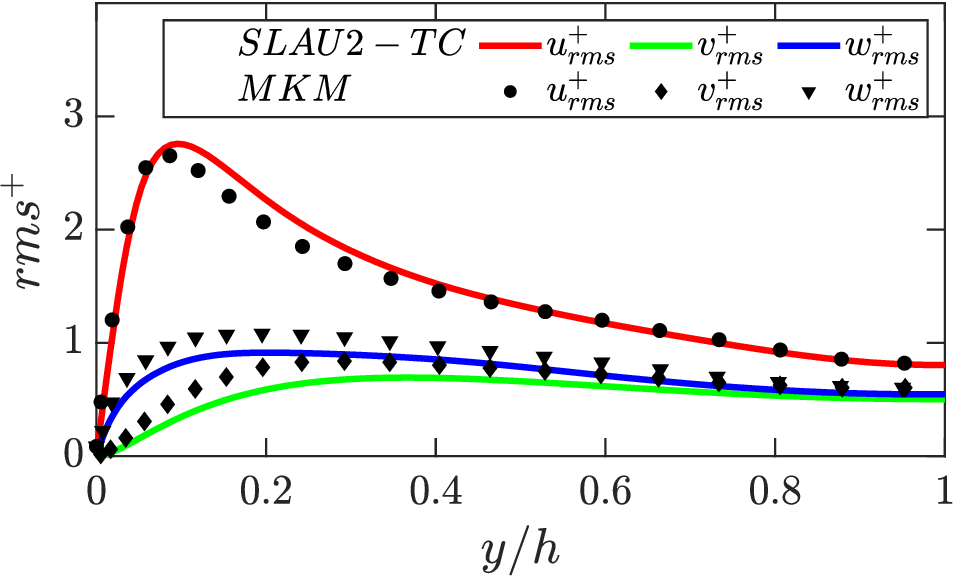}
                \end{minipage}
			\label{fig:rms_val}
		\end{subfigure}
  
		\caption{Variation of (a) mean streamwise velocity and (b) rms of velocity fluctuations along the channel height in a particle-free turbulent channel flow. Results using the upwind SLAU2 scheme with Thornber correction are validated against those by \cite{moser1999direct} (shown using symbols). Reduction in numerical dissipation using the Thornber correction can also be noticed.}
		\label{fig:pfc_stat}
	\end{figure}
    
 Figure \ref{fig:pfc_stat} illustrates the velocity statistics profiles in a PFC flow with $Re_\tau=180$. Once the flow reaches a statistically steady state, 600 samples of flow data at various time instants are collected to calculate the mean flow statistics. The time instants of each sample differed from the previous one by $\frac{\Delta t_\eta}{10}$. Consequently, averaging is performed over a total time period of $60\Delta t_\eta$ and across the two homogeneous directions. Using the mean values of velocity components, the root-mean-square (r.m.s.) values of velocity fluctuations are calculated. 
 
 Figure \ref{fig:pfc_stat}(a) shows the variation of the fluid mean streamwise velocity $\bar{u}$ along the channel height. The plots include results from simulations using the second-order SLAU2 scheme, both with and without the Thornber correction. The results are validated against the previous DNS results of {\cite{moser1999direct}. The figure indicates that the SLAU2 scheme with the Thornber correction produces velocity profiles similar to those by Moser et al. However, without the correction, the scheme overestimates the averaged velocity in regions away from the wall due to increased dissipation. Thus, it can be concluded that the Thornber correction can significantly reduce the unwanted dissipation inherent in the SLAU2 scheme and is used throughout the present study.
 
 Figure \ref{fig:pfc_stat}(b) shows the variation of velocity r.m.s. values in three directions. For clarity, r.m.s. values from simulations without the correction are not included in the figure. Together with figure \ref{fig:pfc_stat}(a), these results demonstrate the ability of the second-order SLAU2 scheme with the Thornber correction to accurately simulate the flow within the incompressible regime.
 
After the framework for clean flow simulations is established for the channel, particles are added to the flow. Upon reaching the new statistically steady state, the averaged flow statistics are calculated in the same manner as done for the PFC flow. A converged state of the PLC flows can be verified from figure \ref{fig:cf} that plots the temporal variation of the skin friction drag reduction factor $DR = (C_{f0}-C_f)/C_{f0}$ in the PLC flows. Here, $C_{f}$ is the skin friction coefficient, defined as $C_f = \frac{2\tau_w\rho}{\dot{m}_f^2}$, which characterises the skin friction drag, $C_{f0}$ is the skin friction coefficient in the PFC flow and $\dot{m}_f$ represents the total mass flux of the fluid. 
 \begin{figure}
		\begin{subfigure}{0.49\textwidth}
			\begin{minipage}[]{0.05\linewidth}
                    \subcaption{}
                \end{minipage}

                \begin{minipage}[]{1\linewidth}
                    \includegraphics[clip=true, trim = 0.0in 0.0in 0.0in 0.0in,width=0.9\textwidth]{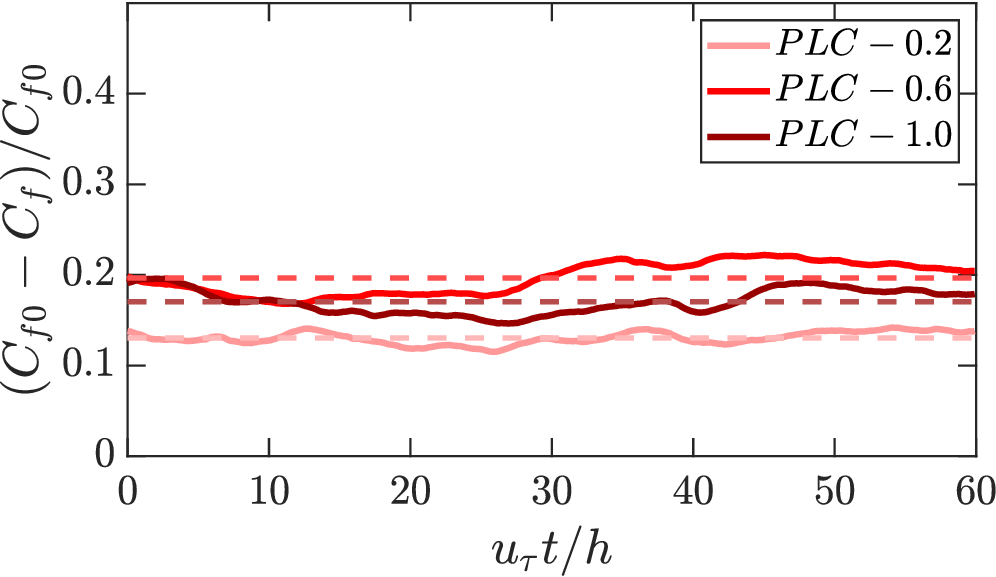}
                \end{minipage}
			\label{fig:cf_p_6}
		\end{subfigure}
		\begin{subfigure}{0.49\textwidth}
			\begin{minipage}[]{0.05\linewidth}
                    \subcaption{}
                \end{minipage}

                \begin{minipage}[]{1\linewidth}
                    \includegraphics[clip=true, trim = 0.0in 0.0in 0.0in 0.0in,width=0.9\textwidth]{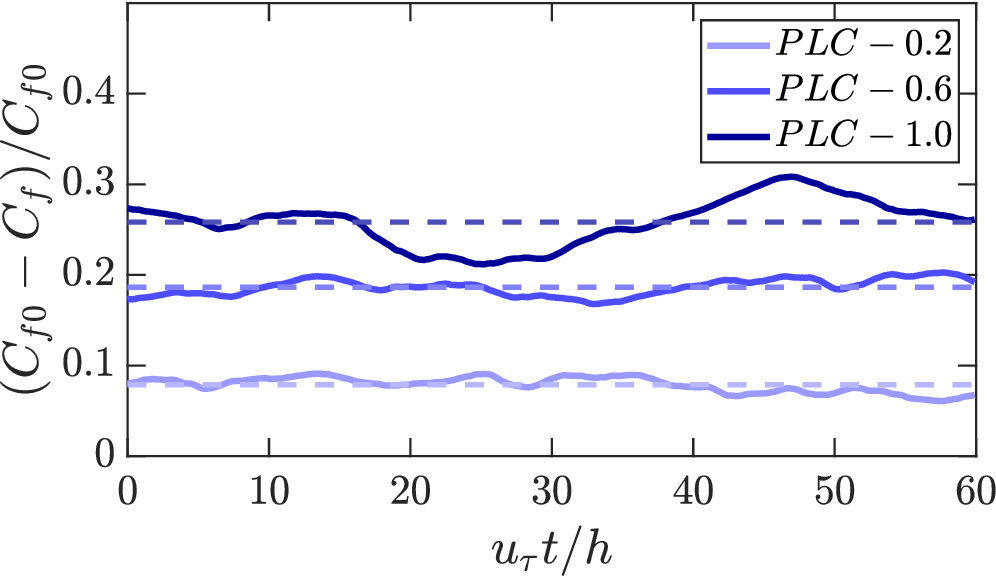}
                \end{minipage}
			\label{fig:cf_p_30}
		\end{subfigure}
        \caption{Temporal variation of the skin friction drag reduction factor $DR = (C_{f0}-C_f)/C_{f0}$ for PLC flows with (a) $St^+ = 6$ and (b) $St^+ = 30$. Darker curves correspond to higher particle mass loadings varying from $0.2$ to $1.0$. Dashed lines represent the time-averaged values of corresponding $DR$.}
		\label{fig:cf}
	\end{figure}

The effect of particles on the fluid mean streamwise velocity $\bar{u}$ is presented in figure \ref{fig:plc_avg} for two different Stokes numbers and three different particle mass loadings for each Stokes number. A validation against the Lagrangian DNS of the PLC flow, under a similar condition of $St^+ = 30$ and $\phi_m=1.0$, carried out by \cite{zhao2013interphasial}, is also demonstrated in the figure. It can be seen that the present Eulerian approach effectively predicts the fluid mean velocity. 
        \begin{figure}
		\begin{subfigure}{0.49\textwidth}
			\begin{minipage}[]{0.05\linewidth}
                    \subcaption{}
                \end{minipage}

                \begin{minipage}[]{1\linewidth}
                    \includegraphics[clip=true, trim = 0.0in 0.0in 0.0in 0.0in,width=0.9\textwidth]{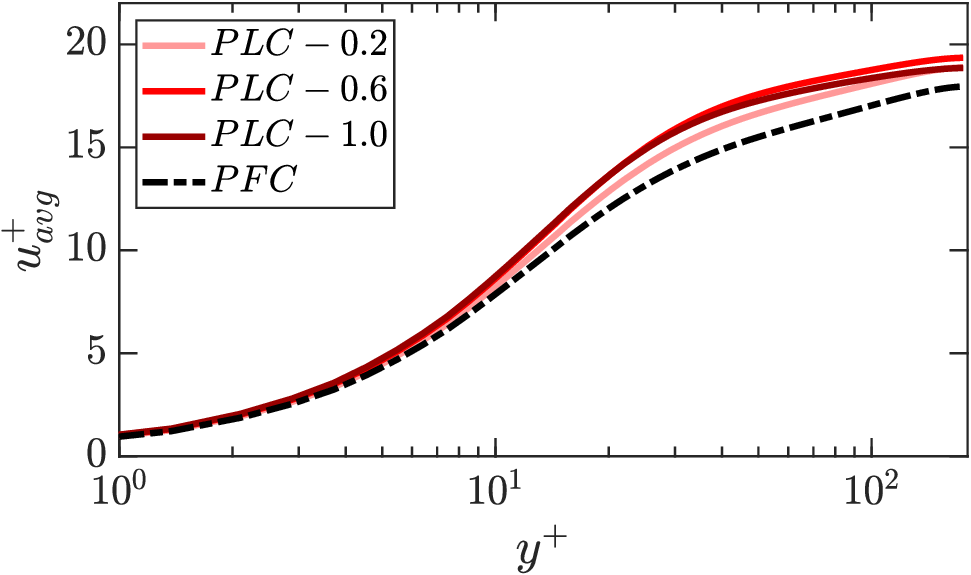}
                \end{minipage}
			\label{fig:avg_p_6}
		\end{subfigure}
		\begin{subfigure}{0.49\textwidth}
			\begin{minipage}[]{0.05\linewidth}
                    \subcaption{}
                \end{minipage}

                \begin{minipage}[]{1\linewidth}
                    \includegraphics[clip=true, trim = 0.0in 0.0in 0.0in 0.0in,width=0.9\textwidth]{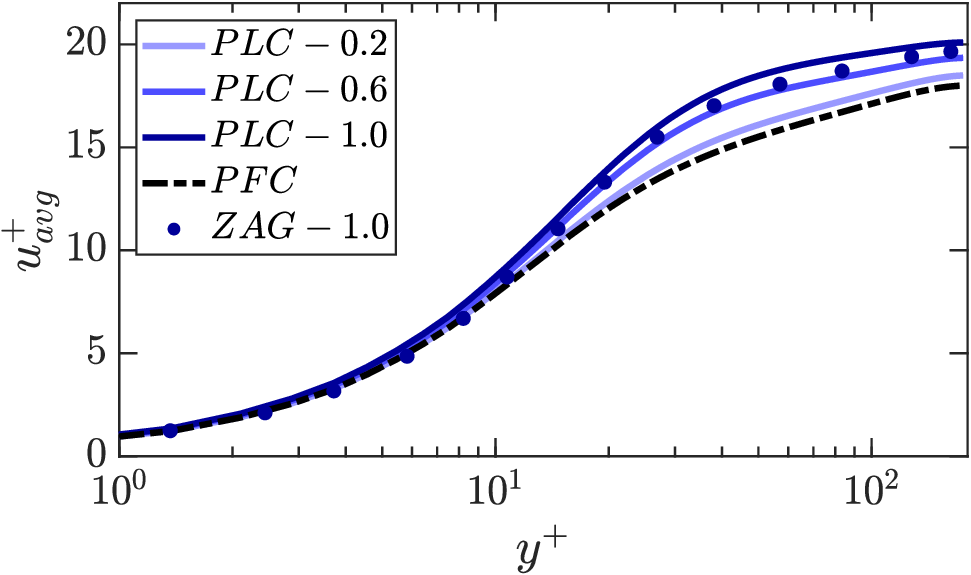}
                \end{minipage}
			\label{fig:avg_p_30}
		\end{subfigure}
        \caption{Variation of fluid mean streamwise velocity along the channel height for the PFC flow and different PLC flows with (a) $St^+ = 6$ and (b) $St^+ = 30$. Darker curves correspond to higher particle mass loadings varying from $0.2$ to $1.0$. The mean flow with $St^+ = 30$ and $\phi_m = 1.0$ is validated against the previous Lagrangian results by \cite{zhao2013interphasial} (ZAG; shown using symbols).}
		\label{fig:plc_avg}
	\end{figure}
 
The figure shows that, in all cases, the presence of particles leads to an increase in the mean streamwise velocity. \cite{dave2023mechanisms} explained that the particles may enhance the flow rate. However, their findings for PLC flows with $St^+ = 6$ are in contrast to the present results. While they assumed the particles with semi-dilute conditions, $\phi_{v0}$ used by them is about one order higher than that used in the present study. Moreover, due to particle clustering, $\phi_v$ can locally increase to a high value \citep{kasbaoui2019clustering}. In the study by \cite{dave2023mechanisms}, it is plausible that particles' $\phi_v$ locally breached the boundary of semi-dilute conditions when inter-particle interactions, which were considered inelastic, become significant.
Furthermore, inelastic collisions were considered between the particles and the wall. It was shown by \cite{vreman2009two} that inelastic inter-particle interactions and particle-wall collisions can flatten the mean velocity profile as the flow loses energy to inelastic interactions. In contrast, the present study, which considers particles in dilute conditions ($\phi_v < 10^{-3}$), involves elastic wall-particle collisions and ignores inter-particle collisions, which may increase the fluid mean velocity.

Though the addition of particles increased $\bar{u}$, this increase in the fluid mean velocity is affected by $\phi_m$. For $St^+ = 6$, $\bar{u}$ increases as $\phi_m$ increases from 0.2 to 0.6. Notably, further increase in $\phi_m$ to 1.0 reduces $\bar{u}$. However, for the flow with $St^+ = 30$, we find that $\bar{u}$ increases monotonically with $\phi_m$. This influence of particles on the fluid mean velocity can be explained using figure \ref{fig:cf}.  For $St^+ = 6$, the average value of $DR$ increases when $\phi_m$ is increased from $0.2$ to $0.6$, but it reduces with a further increase in $\phi_m$ to $1.0$. For $St^+ = 30$, the average value of $DR$ shows a monotonic increase with $\phi_m$. Thus, it can be concluded that, depending on the Stokes number, increasing the particle mass loading beyond a certain limit may increase the skin friction drag, as a result of which the fluid mean velocity may start to decrease.

\begin{figure}
            \begin{subfigure}{0.49\textwidth}
			\begin{minipage}[]{0.05\linewidth}
                    \subcaption{}
                \end{minipage}

                \begin{minipage}[]{1\linewidth}
                    \includegraphics[clip=true, trim = 0.0in 0.0in 0.0in 0.0in,width=0.9\textwidth]{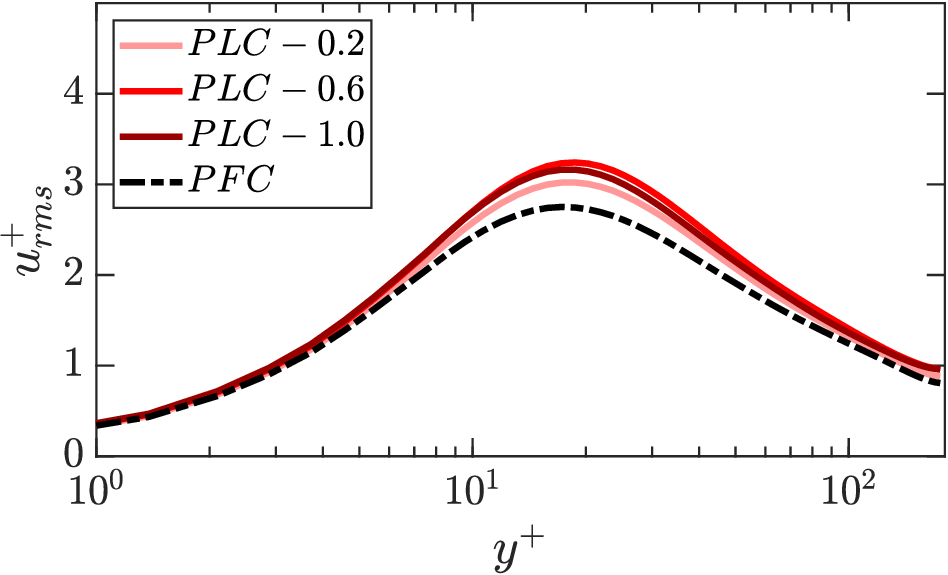}
                \end{minipage}
			\label{fig:urms_p_6}
		\end{subfigure}
		\begin{subfigure}{0.49\textwidth}
			\begin{minipage}[]{0.05\linewidth}
                    \subcaption{}
                \end{minipage}

                \begin{minipage}[]{1\linewidth}
                    \includegraphics[clip=true, trim = 0.0in 0.0in 0.0in 0.0in,width=0.9\textwidth]{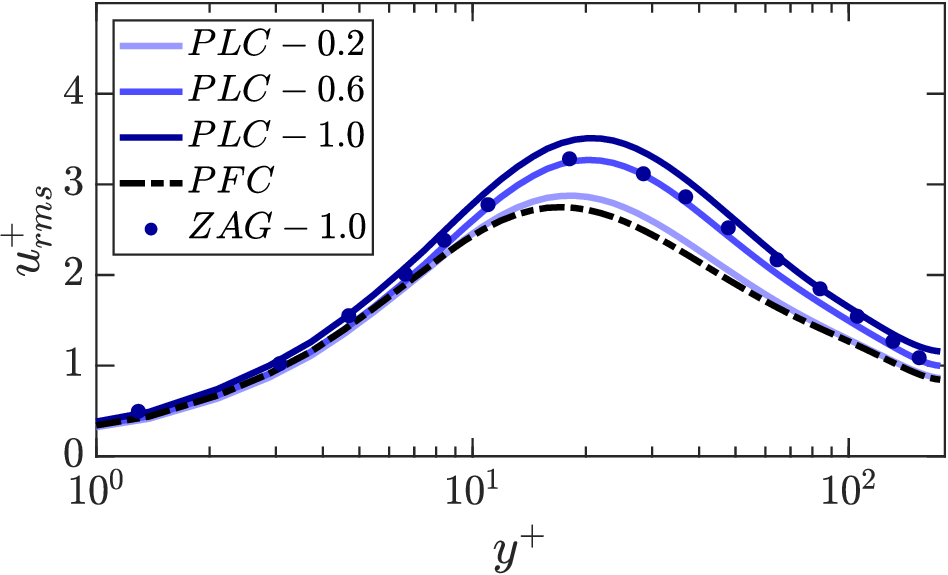}
                \end{minipage}
			\label{fig:urms_p_30}
		\end{subfigure}

             \begin{subfigure}{0.49\textwidth}
			\begin{minipage}[]{0.05\linewidth}
                    \subcaption{}
                \end{minipage}

                \begin{minipage}[]{1\linewidth}
                    \includegraphics[clip=true, trim = 0.0in 0.0in 0.0in 0.0in,width=0.9\textwidth]{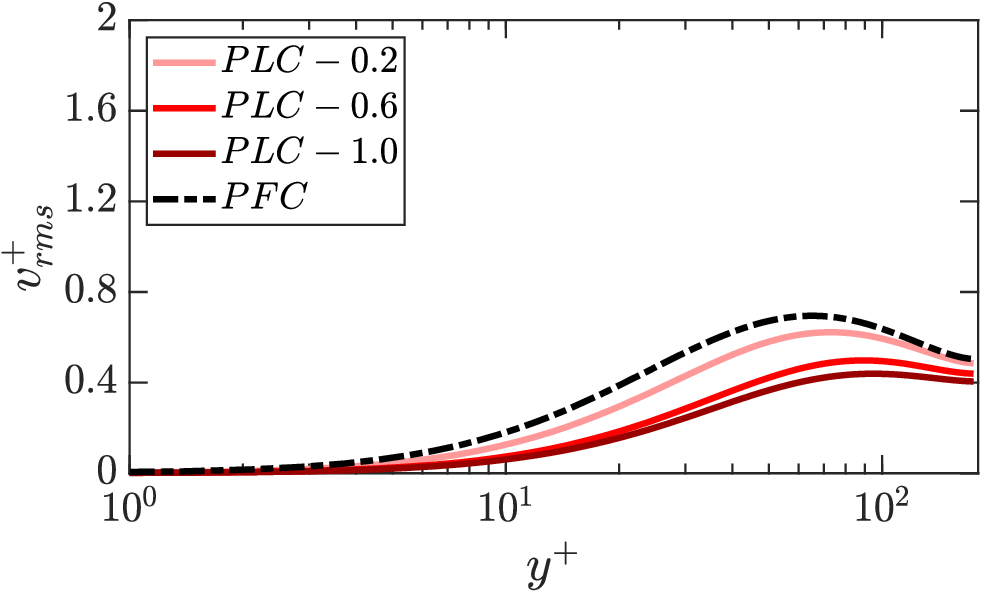}
                \end{minipage}
			\label{fig:vrms_p_6}
		\end{subfigure}
		\begin{subfigure}{0.49\textwidth}
			\begin{minipage}[]{0.05\linewidth}
                    \subcaption{}
                \end{minipage}

                \begin{minipage}[]{1\linewidth}
                    \includegraphics[clip=true, trim = 0.0in 0.0in 0.0in 0.0in,width=0.9\textwidth]{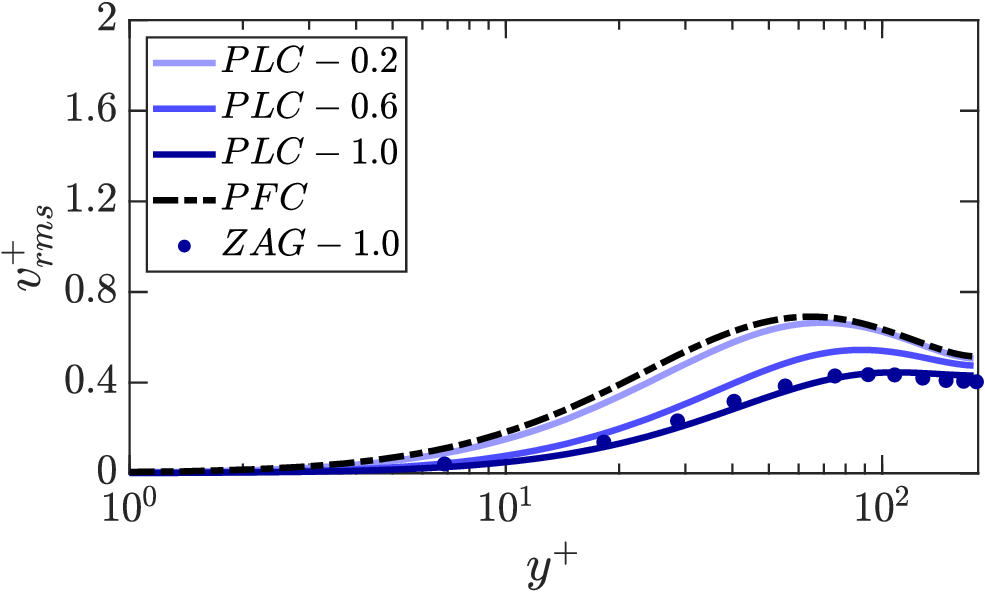}
                \end{minipage}
			\label{fig:vrms_p_30}
		\end{subfigure}

             \begin{subfigure}{0.49\textwidth}
			\begin{minipage}[]{0.05\linewidth}
                    \subcaption{}
                \end{minipage}

                \begin{minipage}[]{1\linewidth}
                    \includegraphics[clip=true, trim = 0.0in 0.0in 0.0in 0.0in,width=0.9\textwidth]{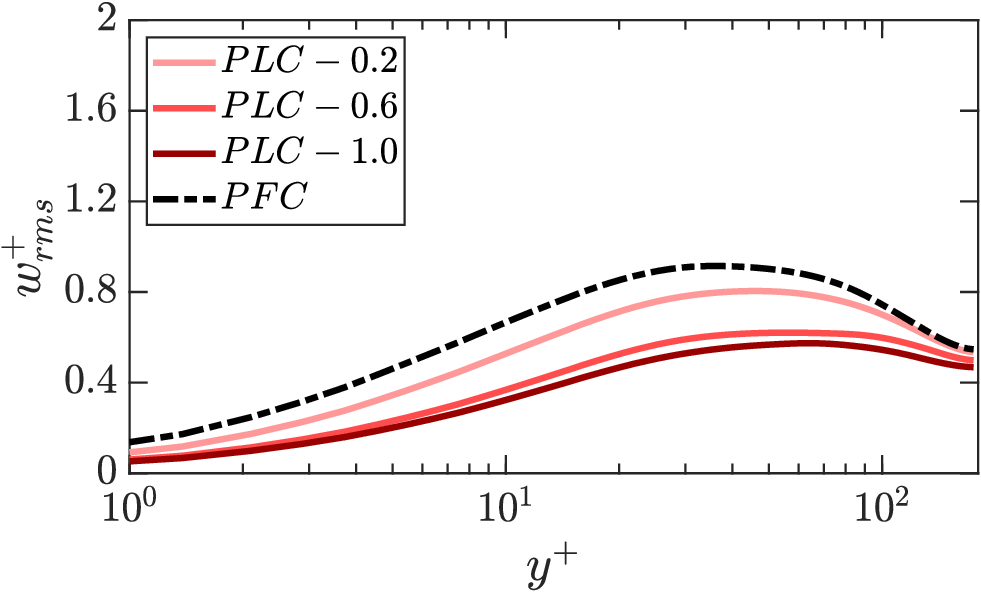}
                \end{minipage}
			\label{fig:wrms_p_6}
		\end{subfigure}
		\begin{subfigure}{0.49\textwidth}
			\begin{minipage}[]{0.05\linewidth}
                    \subcaption{}
                \end{minipage}

                \begin{minipage}[]{1\linewidth}
                    \includegraphics[clip=true, trim = 0.0in 0.0in 0.0in 0.0in,width=0.9\textwidth]{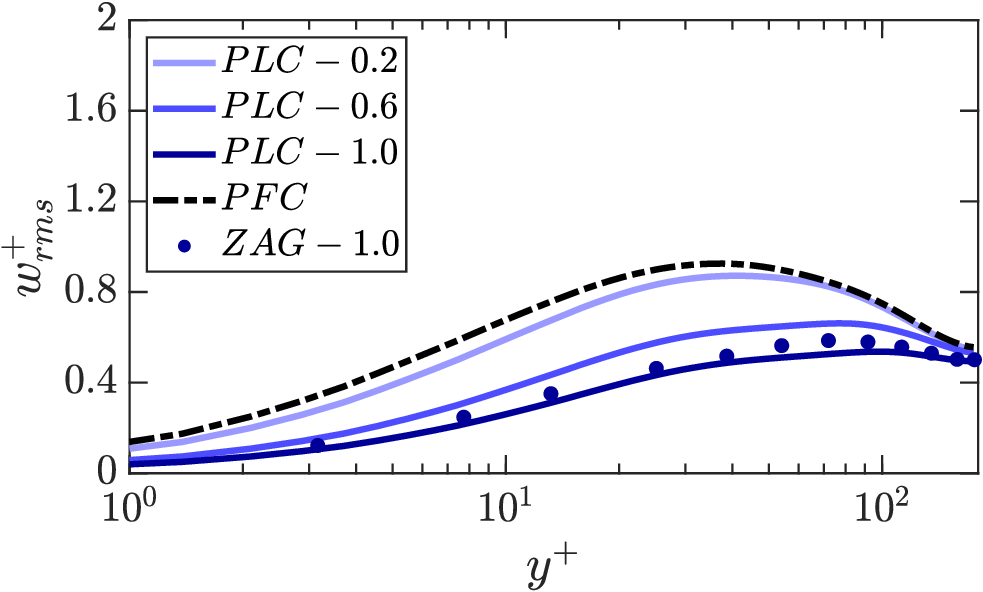}
                \end{minipage}
			\label{fig:wrms_p_30}
		\end{subfigure}
  
		\caption{Variation of fluid r.m.s. velocities along the channel height for the PFC flow and different PLC flows with (a, c, e) $St^+ = 6$ and (b, d, f) $St^+ = 30$. Darker curves correspond to higher particle mass loadings varying from $0.2$ to $1.0$. The r.m.s. values for the flow with $St^+ = 30$ and $\phi_m = 1.0$ are validated against the values obtained using the Lagrangian simulation by \cite{zhao2013interphasial}
        (ZAG; shown using symbols).}
		\label{fig:plc_fluc}
	\end{figure} 
The influence of particles on fluid velocity fluctuations is effectively captured using the Eulerian approach, as shown in figure \ref{fig:plc_fluc}. The figure establishes the credibility of the present Eulerian approach in predicting the fluctuations by validating the results for $St^+ = 30$ and $\phi_m = 1.0$ case against the Lagrangian DNS of similar PLC flow carried out by \cite{zhao2013interphasial}. Although there is a slight difference between the two streamwise fluctuations near $y^+ = 30$, the maximum relative difference is around $6\%$. It is well established that the main influence of particles on turbulence modulation may be parametrised by the Stokes number and the particle mass loading \citep{kulick1994particle,zhao2010turbulence,zhao2013interphasial,dave2023mechanisms}.
Here, we test this hypothesis using the Eulerian approach. It should be mentioned that to always keep the volume fraction low enough to ensure a dilute mixture, including in regions of particle clustering, it was not possible to compare our methodology with previous Euler-Lagrangian studies with the exact same flow conditions. 
Markedly, although the volume fraction differs from that of the work of \cite{zhao2013interphasial}, it is shown that both flow statistics and mean drag (figure \ref{fig:drag_force}) are similar.

From the figure, it can be inferred that particles enhance streamwise fluctuations and attenuate transverse ones. \cite{kulick1994particle} reported and explained this anisotropic effect of particles on velocity fluctuations. The high frequencies of transverse fluctuations make it difficult for inertial particles to adjust to the flow conditions. Thus, the turbulent energy of these fluctuations is dissipated by the particles. With an increase in the number of particles, more energy is dissipated in the transverse directions. On the other hand, particles adjust to streamwise fluctuations that are at relatively lower frequencies and hence do not dissipate the turbulent energy. Instead, particles enhance fluctuations in the streamwise direction. However, the current results show that this enhancement in streamwise velocity fluctuations is affected by the particle $\phi_m$, in the same way as the fluid streamwise mean velocity.

\subsection{\label{subsec:parclus}Near-wall particle accumulation and preferential clustering}
    \begin{figure}
        \includegraphics[clip=true, trim = 0.2in 1.3in 0.1in 0.2in,width=0.96\textwidth]{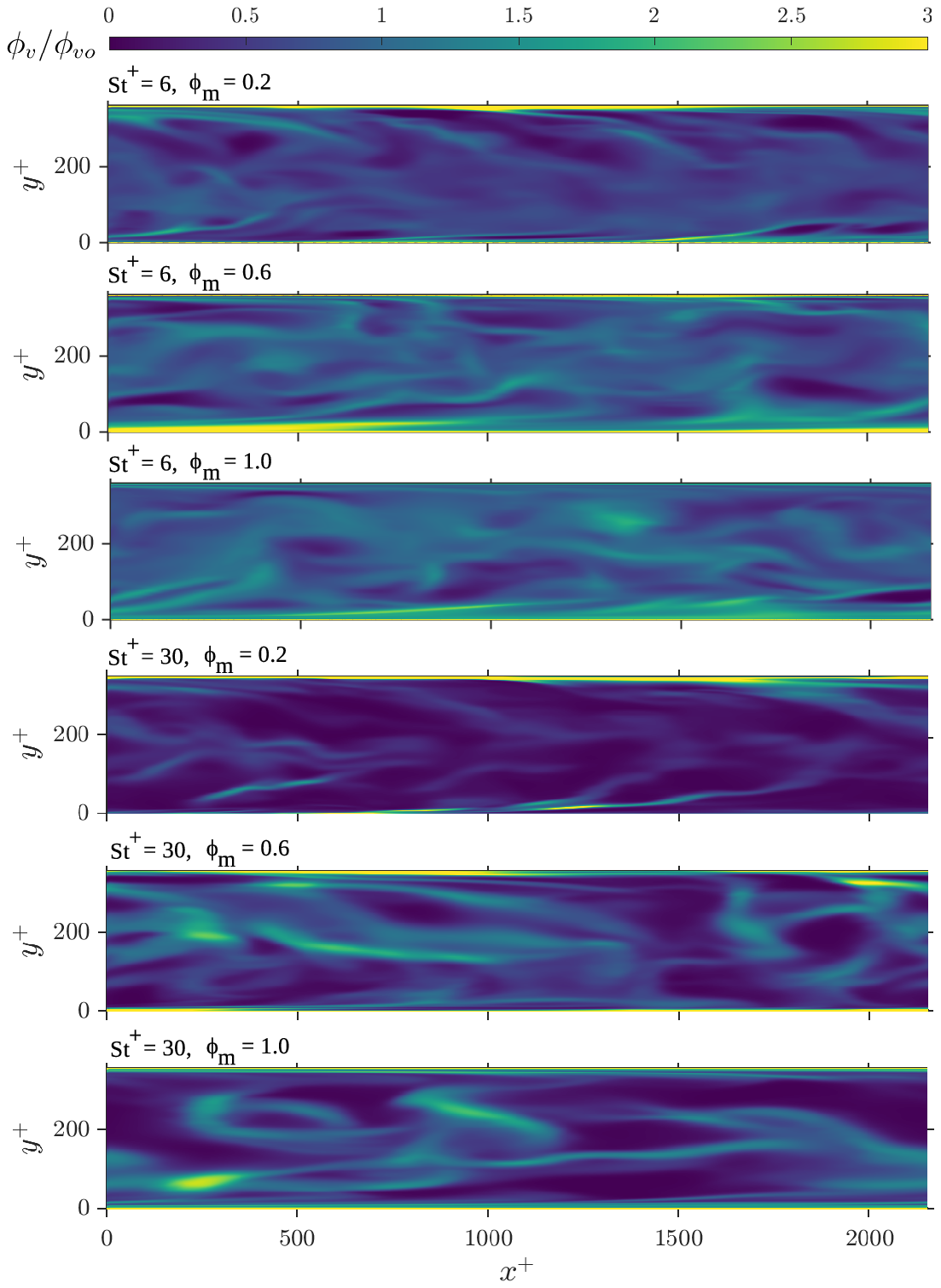}
            \caption{Contour plots of instantaneous particle volume fraction, normalised by the initial volume fraction, in the wall-normal plane, for different combinations of $St^+$ and $\phi_m$. The plots show particle streaks aligned in the streamwise direction. High values of $\phi_v/\phi_{v0}$ in the near-wall regions indicate particle accumulation.}
		\label{fig:phi_p_contour}
        \end{figure}
        \quad Particle migration towards the wall and their clustering in a wall-normal plane of turbulent channel flow at any time instant can be quantified by comparing the instantaneous particle volume fraction $\phi_v$ with their initial volume fraction $\phi_{v0}$ when the particle distribution is uniform throughout the channel. This is demonstrated in figure \ref{fig:phi_p_contour} that shows instantaneous contour plots of $\phi_v$, normalised by $\phi_{v0}$, in a wall-normal plane at the channel centre. All subfigures demonstrate the capability of the present Eulerian method to capture the intricate particle field, showing the streaks of particles elongated in the streamwise direction. Streaks in flows with $St^+ = 30$ are found to be longer and more intense than in flows with $St^+ = 6$. In all the cases, particles' tendency to migrate and accumulate near the wall can be noticed. 
        
        This phenomenon is demonstrated in figure \ref{fig:phi_p}, which shows the variation of the time-averaged values of the normalised $\phi_v$ along the channel height. It can be seen that, relative to the initial volume fraction, the particle accumulation at the wall decreases with an increase in $\phi_m$. As a result of the conservation of the total number of particles in the channel, the void density near the channel centre is higher at lower values of $\phi_m$. Although particles' tendency to migrate towards the wall decreases with an increase in $\phi_m$, our data show that the average volume fraction $\overline{\phi}_v$ of particles at any location is higher for higher $\phi_m$. High-inertia particles in PLC flows with $St^+ = 30$ tend to have a greater near-wall accumulation than low-inertia particles in PLC flows with $St^+ = 6$. As illustrated in table \ref{tab:param}, at a particular $\phi_m$, the initial volume fraction is similar for each particle type. This means the average number density of the low-inertia particles (with a lower particle volume) is higher than that of the high-inertia particles (with a higher particle volume). Consequently, flows with $St^+ = 30$ exhibit more voids in the channel core region than flows with $St^+ = 6$. These observations are similar to those obtained using the Lagrangian approach \citep{dave2023mechanisms}. Previous studies have attributed this phenomenon of particle accumulation near the wall to the particles' inertia, due to which the particles migrate to low-turbulence regions by the effect of turbophoresis \citep{reeks1983transport,kuerten2011turbulence,nowbahar2013turbophoresis,johnson2020turbophoresis}.
        
        Besides migration to the wall, particles can also preferentially cluster along the velocity streaks. In a turbulent channel flow, the phenomenon of preferential clustering in LSS was found to be more dominant for smaller dense particles \citep{eaton1994preferential,fong2019velocity,dave2023mechanisms}. In comparison, larger buoyant particles prefer HSS \citep{zhu2020interface,peng2024preferential}. Using the present Eulerian approach for the particle field, we found the preferential clustering of particles in LSS for all cases.

        \begin{figure}
		\begin{subfigure}{0.49\textwidth}
                \begin{minipage}[]{0.05\linewidth}
                    \subcaption{}
                \end{minipage}

                \begin{minipage}[]{1\linewidth}
			     \includegraphics[clip=true, trim = 0.5in 3.4in 0.4in 3.6in,width=0.9\textwidth]{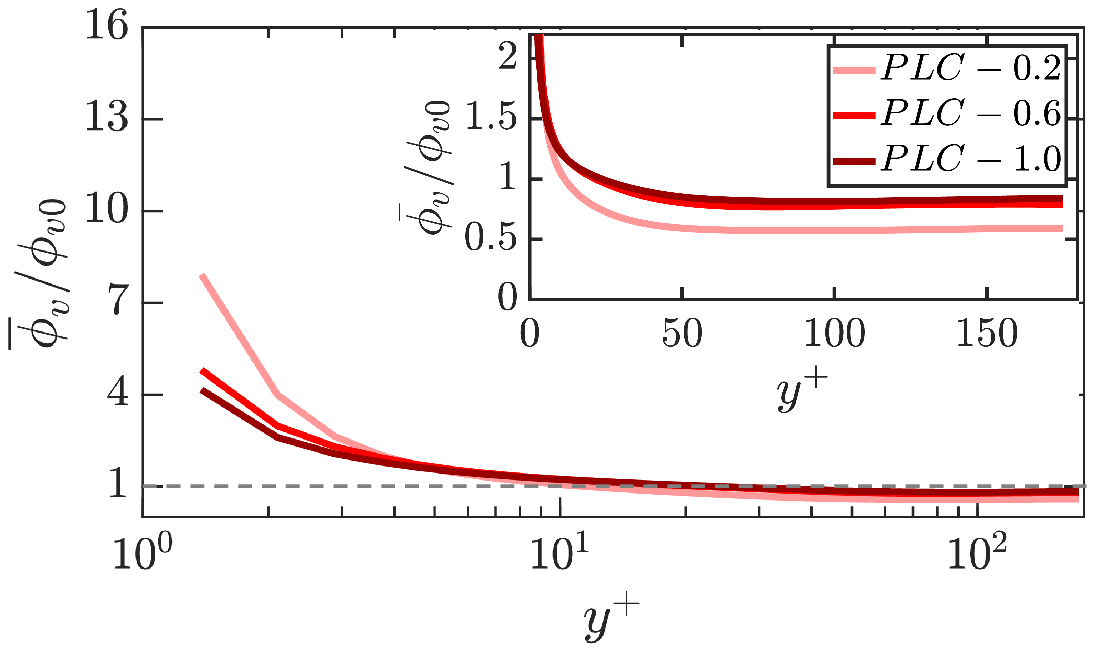}
                \end{minipage}
			\label{fig:phi_p_6}
		\end{subfigure}
		\begin{subfigure}{0.49\textwidth}
                \begin{minipage}[]{0.05\linewidth}
                    \subcaption{}
                \end{minipage}

                \begin{minipage}[]{1\linewidth}
                   \includegraphics[clip=true, trim = 0.5in 3.4in 0.4in 3.6in,width=0.9\textwidth]{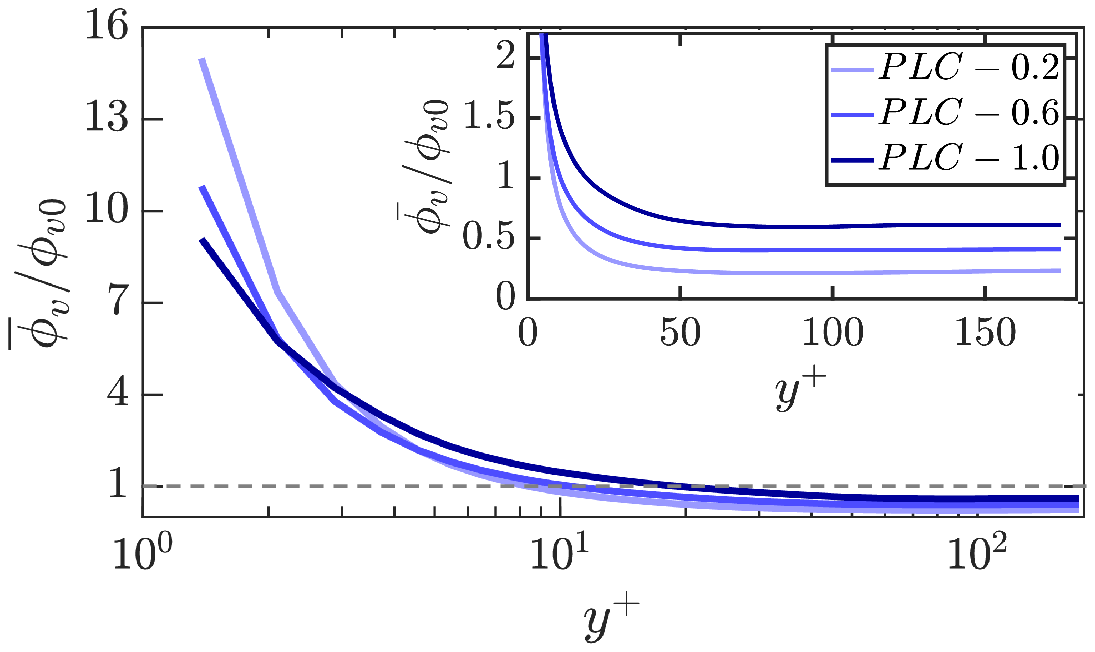} 
                \end{minipage}
			\label{fig:phi_p_30}
		\end{subfigure}
  
		\caption{Variation of mean particle volume fraction along the channel height for different PLC flows with (a) $St^+ = 6$ and (b) $St^+ = 30$. The inset compares the particle volume fraction near the channel centre for different particle mass loadings. The dashed line represents the reference initial particle volume fraction. Darker curves correspond to higher particle mass loadings varying from $0.2$ to $1.0$.}
		\label{fig:phi_p}
	\end{figure}
 
     \begin{figure}
		\begin{subfigure}{0.49\textwidth}
                \begin{minipage}[]{0.05\linewidth}
                    \subcaption{}
                \end{minipage}

                \begin{minipage}[]{1\linewidth}
                    \includegraphics[clip=true, trim = 0.0in 0.0in 0.0in 0.0in,width=0.9\textwidth]{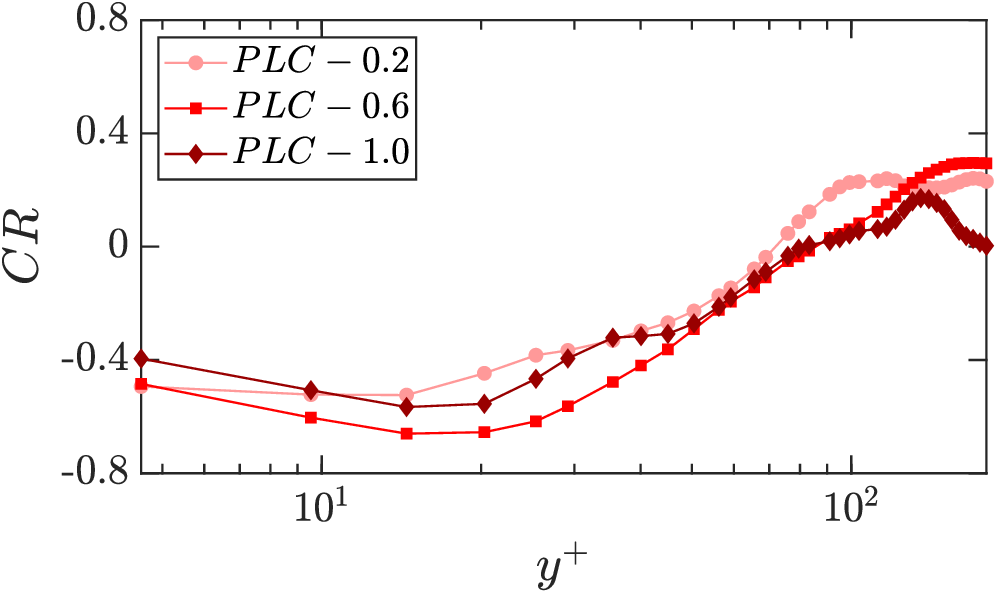}
                \end{minipage}
			\label{fig:corr_p_6}
		\end{subfigure}
		\begin{subfigure}{0.49\textwidth}
		    \begin{minipage}[]{0.05\linewidth}
                    \subcaption{}
                \end{minipage}
                
                \begin{minipage}[]{1\linewidth}
                    \includegraphics[clip=true, trim = 0.0in 0.0in 0.0in 0.0in,width=0.9\textwidth]{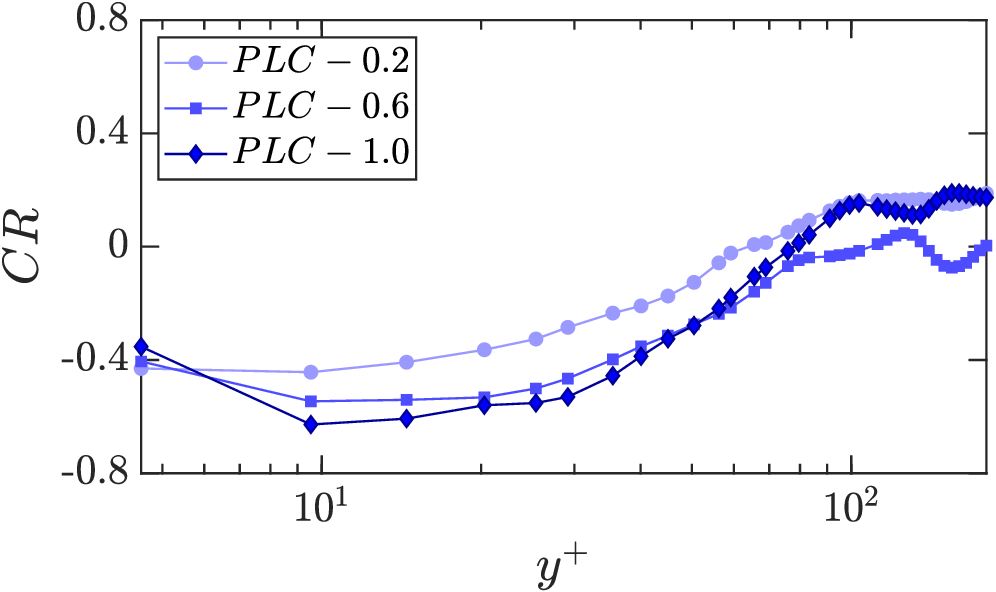}
                \end{minipage}
			\label{fig:corr_p_30}
		\end{subfigure}
  
		\caption{Pearson correlation coefficient between the instantaneous particle volume fraction and the instantaneous fluid streamwise velocity fluctuations in spanwise planes at different $y^+$ for different PLC flows with (a) $St^+ = 6$ and (b) $St^+ = 30$. Darker curves correspond to higher particle mass loadings varying from $0.2$ to $1.0$.}
		\label{fig:corr_p}
	\end{figure}   

    \begin{figure}
		\begin{subfigure}{0.49\textwidth}
                \begin{minipage}[]{0.05\linewidth}
                    \subcaption{}
                \end{minipage}

                \begin{minipage}[]{1\linewidth}
                    \includegraphics[clip=true, trim = 0.4in 3.5in 0.8in 3.8in,width=0.9\textwidth]{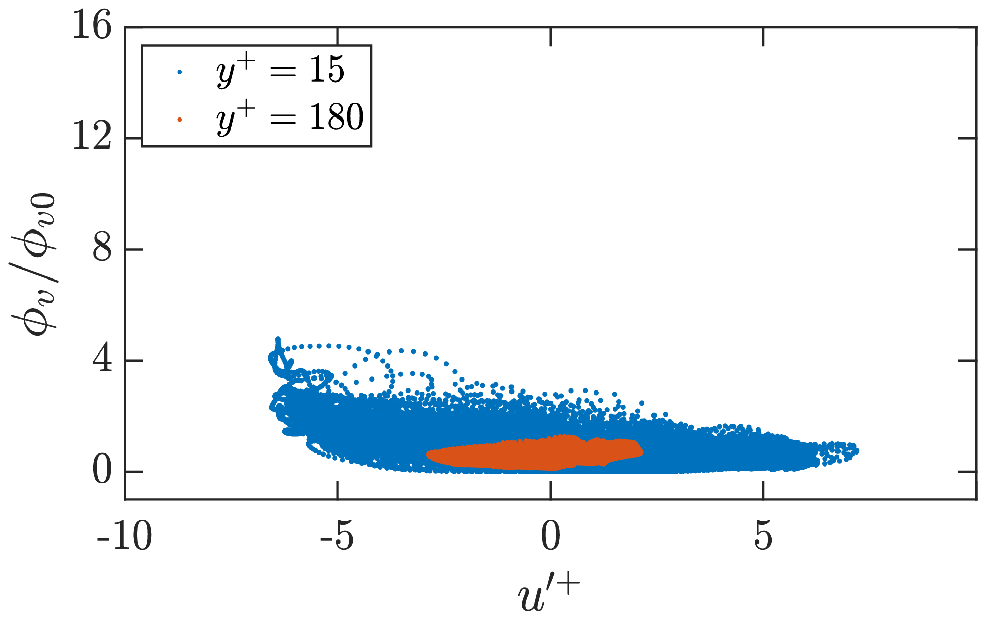}
                \end{minipage}
			\label{fig:ro_p_u_fluc_p_6}
		\end{subfigure}
		\begin{subfigure}{0.49\textwidth}
                \begin{minipage}[]{0.05\linewidth}
                    \subcaption{}
                \end{minipage}

                \begin{minipage}[]{1\linewidth}
                    \includegraphics[clip=true, trim = 0.4in 3.5in 0.8in 3.8in,width=0.9\textwidth]{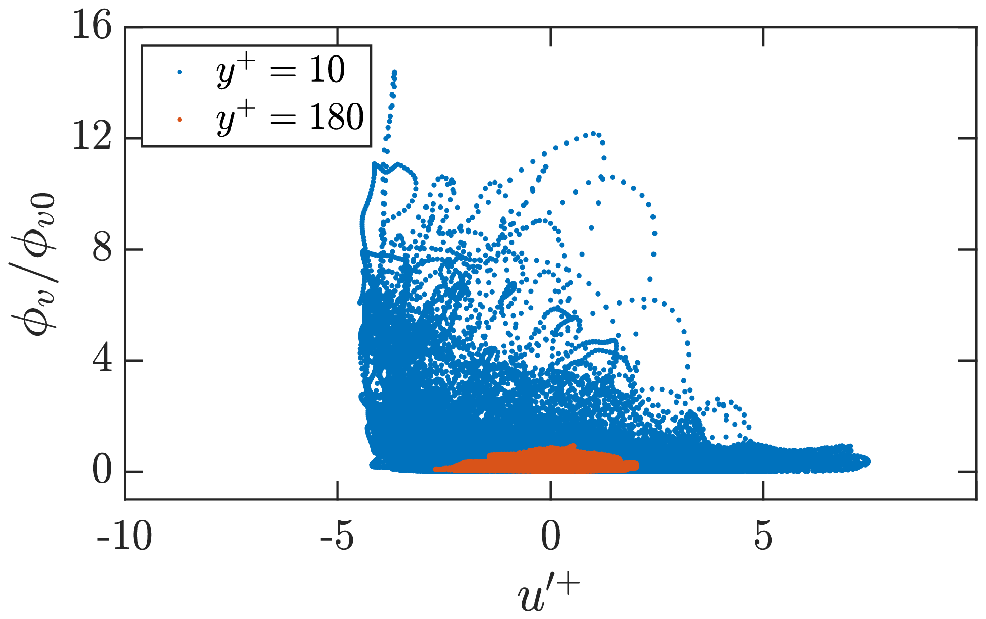}
                \end{minipage}
			\label{fig:ro_p_u_fluc_p_30}
		\end{subfigure}
  
		\caption{Normalised particle volume fraction at different values of the fluid streamwise velocity fluctuations in two planes of the PLC flow with (a) $St^+ = 6$ and (b) $St^+ = 30$. The near-wall plane is at the lower $y^+$ where the correlation shown in figure \ref{fig:corr_p} is most negative, and the plane at the channel centre is at $y^+ \approx 180$. Instantaneous data at the same time instants as in figure \ref{fig:corr_p} is analysed. For both flows $\phi_m = 0.2$.}
		\label{fig:ro_p_u_fluc_p}
	\end{figure}

    Figure \ref{fig:corr_p} plots the linear correlation between the instantaneous particle volume fraction and the streamwise velocity fluctuations in various spanwise planes along the channel height, calculated using the Pearson correlation coefficient. For both types of particles, a negative correlation is observed near the wall. This signifies that particles near the wall prefer to cluster in the region of LSS. However, it is difficult to effectively predict the correlation away from the wall due to the very low particle volume fraction there. 
    
    For $St^+ = 6$, the best negative correlation is obtained near $Y^+ = 15$, while for $St^+ = 30$, it is obtained near $Y^+ = 10$. At the same time instants, figure \ref{fig:ro_p_u_fluc_p} maps the particle volume fraction $\phi_v$ to the fluid streamwise velocity fluctuations in the near-wall spanwise plane with the most negative correlation and compares it with that in the plane away from the wall ($y^+ \approx 180$), at the lowest particle mass loading $\phi_m = 0.2$. It can be seen that for both types of particles, preferential clustering is significant in the near-wall region, with more particles mapping the negative velocity fluctuations than the positive ones. Away from the wall, the particles do not show any significant preference towards any type of velocity fluctuations. Thus, it can be concluded that the preferential clustering of particles is a near-wall phenomenon. 
    
    Near the wall, different scales of LSS can be present. Particles' preference towards different scales of LSS can be explained using the clustering probability calculated for different sets of velocity streaks. A velocity streak set contains the bounded values of the fluid streamwise velocity fluctuations $u'^+$. If a specific location in the plane has a velocity streak corresponding to a particular set, the clustering probability determines the probability of particle clustering at that location for the corresponding set. We consider clustering at a particular location when $\phi_v$ at that location is greater than the planar average $\bar{\phi}_{vp}$. Thus, the clustering probability for a particular streak set $V$ at a location in the plane is calculated as:
    \begin{equation}
    P[(\phi_v>\bar{\phi}_{vp})|(u'\in V)]= \frac{P[(\phi_v>\bar{\phi}_{vp})\cap(u'\in V)]}{P[u'\in V]}
    \end{equation}

    At the same time instants, as in figure \ref{fig:corr_p}, we find that the instantaneous $u'^+$ varies from -8 to 8. These fluctuations are divided into eight streak sets, with each being determined by its mean value,
    $
        [-8,-6],[-6,-4],[-4,-2],[-2,0],[0,2],[2,4],[4,6],[6,8].
    $
    The calculated clustering probabilities in the near-wall plane are presented in figure \ref{fig:probro_p_u_fluc}. It is clear that for both types of particles for different mass loadings, clustering is strongest in the most negative streaks. In the regions of weak LSS, the probability of particle clustering decreases. The probability of finding particle clusters in HSS is relatively very low. Although particle clustering in LSS is observed in all the cases, the comparison of clustering intensity in different cases illustrates a complex picture.
    \begin{figure}
		\begin{subfigure}{0.49\textwidth}
                \begin{minipage}[]{0.05\linewidth}
                    \subcaption{}
                \end{minipage}

                \begin{minipage}[]{1\linewidth}
                    \includegraphics[clip=true, trim = 0in 0.0in 0in 0in,width=0.9\textwidth]{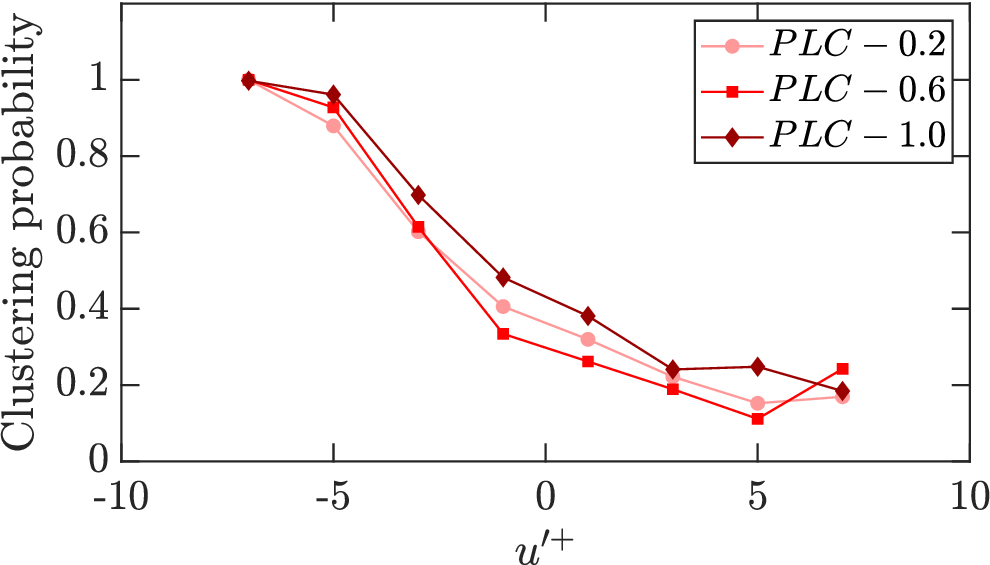}
                \end{minipage}
			\label{fig:probro_p_u_fluc_6}
		\end{subfigure}
		\begin{subfigure}{0.49\textwidth}
                \begin{minipage}[]{0.05\linewidth}
                    \subcaption{}
                \end{minipage}

                \begin{minipage}[]{1\linewidth}
                    \includegraphics[clip=true, trim = 0in 0.0in 0in 0in,width=0.9\textwidth]{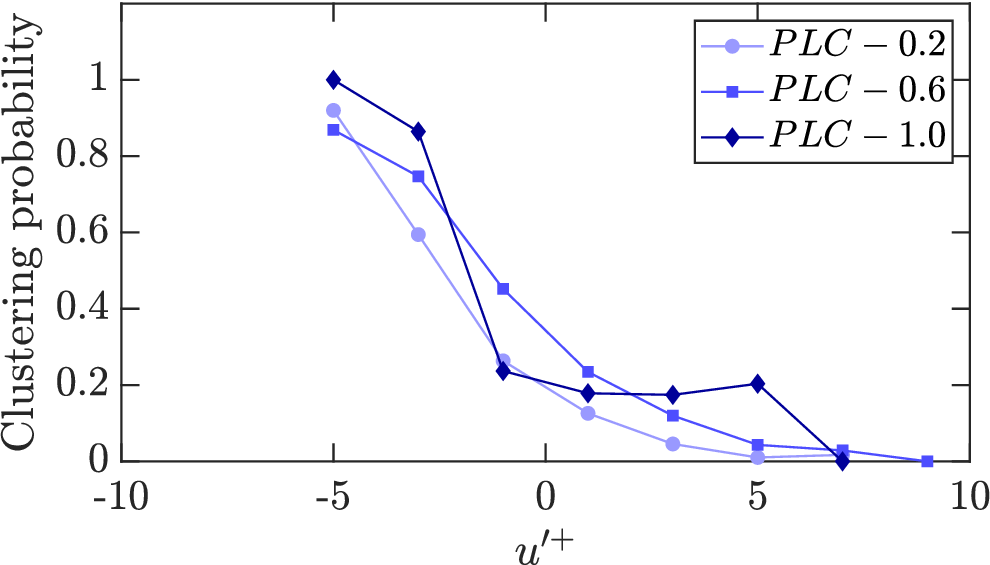}
                \end{minipage}
			\label{fig:probro_p_u_fluc_30}
		\end{subfigure}
  
		\caption{Instantaneous particle clustering probability in the near-wall plane for different values of streamwise velocity fluctuations in PLC flows with (a) $St^+ = 6$ where the considered near-wall plane is at $y^+ \approx 15$ and (b) $St^+ = 30$ where the considered near-wall plane is at $y^+ \approx 10$. Instantaneous data at the same time instants as in figure \ref{fig:corr_p} are analysed. Darker symbols correspond to higher particle mass loadings varying from $0.2$ to $1.0$.}
		\label{fig:probro_p_u_fluc}
	\end{figure}

    \begin{table}
        \centering
        \begin{tabular}{p{0.07\linewidth} p{0.12\linewidth} p{0.12\linewidth} p{0.12\linewidth}}
           $St^+$  &  $\phi_m = 0.2$ & $\phi_m = 0.6$ & $\phi_m = 1.0$ \\
           & & & \\
           $6$  & $0.4064$ & $0.3747$ & $0.4097$ \\
           $30$ & $0.2351$ & $0.3238$ & $0.3585$ \\
        \end{tabular}
        \caption{Instantaneous probability of finding the particle clusters in the spanwise plane, at $y^+ \approx 15$ for PLC flows with $St^+ = 6$ and at $y^+ \approx 10$ for PLC flows with $St^+ = 30$. Instantaneous data at the same time instants as in figure \ref{fig:corr_p} are analysed. Particle clusters are considered at any location when the local $\phi_v > \bar{\phi}_{vp}$.}
        \label{tab:parclus}
    \end{table}
    It may happen that for a particular case, the value of $\phi_v$ is high at certain locations, but the other case may have more locations with a lower value of $\phi_v$ (but greater than $\bar{\phi}_{vp}$). By comparing the fraction of the planar area at which $\phi_v > \bar{\phi}_{vp}$, an attempt has been made to compare particle clustering for different cases. This gives us the probability of finding the particle clusters in the area. The probability values are tabulated in table \ref{tab:parclus}. The table shows that for $St^+ = 30$, the probability of finding particle clusters increases with $\phi_m$, while for $St^+ = 6$, $\phi_m$ has a less effect on the clustering of particles. However, the probability for $St^+ = 6$ is always higher than that for $St^+ = 30$. Thus, it can be concluded that the low-inertia particles exhibit preferential clustering more than the high-inertia particles. This is consistent with the notion that the high-inertia particles are less responsive to velocity fluctuations and thus exhibit less clustering. Here, it is worth mentioning that \cite{sardina2012wall}, through their simulation data of one-way coupled particle-laden wall flows, showed that the smaller domain may cause phase locking of the largest-velocity structures. This can lead to some inaccuracies in the prediction of particle accumulation and particle preferential clustering. However, in a two-way coupled flow, the particles can cause the redistribution of turbulent energy to the smaller structures, and this redistribution increases with the particle mass loading \citep{kulick1994particle}. Moreover, \cite{sardina2012wall} pointed out that single-point correlations were unaffected by the domain truncation. For our simulation, we have considered the largest domain that was used in the previous point-particle Lagrangian simulations of two-way coupled PLC flows with elastic wall collisions \citep{zhao2010turbulence,zhao2013interphasial,zhou2020non,dave2023mechanisms}. For the present study that aims to capture various physics related to particle-flow interactions using an Eulerian approach, the choice of the domains used is adequate.

\subsection{\label{subsec:accu_eff}Effects of particle accumulation and preferential clustering on flow dynamics}
\quad Particle accumulation and preferential clustering have a significant effect on various fluid dynamics, including the particle mass flow rate, totalRSS, interphase drag, and turbulence modulation. In this section, these phenomena are investigated.

\subsubsection{\label{subsubsec:parmass}Particle mass flow rate}
        
    \quad For incompressible flows, the effect of particles on the fluid mass flow rate through the channel can be estimated to be the same as the fluid mean streamwise velocity. The mass flow rate of particles $\dot{m}_p$ through the channel can be greatly affected by $\phi_m$ and $St^+$ as shown in figure \ref{fig:m_dot_p} that plots the average mass flow rate, normalised by the total fluid mass flow of the PFC flow ($\tilde{\dot{m}}_p = \overline{\dot{m}}_p/m_0$), per unit $\phi_m$ of the respective case. Referring to figure \ref{fig:phi_p}, it can be inferred that the particle mass flow rate is a direct consequence of particle migration towards the walls. Due to slip and reflective conditions, $\dot{m}_p$ at the wall is not zero. Furthermore, it can be observed that the lowest value of $\tilde{\dot{m}}_p$ is not located at the wall but in a region slightly away from the wall. At the same $St^+$, although the average mass flow rate increases throughout the channel height with increasing $\phi_m$, $\tilde{\dot{m}}_p$ represents a contrasting picture near the wall. This is due to the decreased tendency of particles to migrate to the wall with increasing $\phi_m$. The figure also shows that the effect of the Stokes number on $\tilde{\dot{m}}_p$ in the near-wall region is in contrast to its effect in the region away from the wall; while $\tilde{\dot{m}}_p$ in the near-wall region increases with an increase in $St^+$, $\tilde{\dot{m}}_p$ in the region away from the wall decreases. At higher $St^+$, increased particle accumulation at the wall causes an increase in $\tilde{\dot{m}}_p$, whereas increased particle void density in the channel core region leads to a decrease in $\tilde{\dot{m}}_p$.
    \begin{figure}
		\begin{subfigure}{0.49\textwidth}
			\begin{minipage}[]{0.05\linewidth}
                    \subcaption{}
                \end{minipage}

                \begin{minipage}[]{1\linewidth}
                    \includegraphics[clip=true, trim = 0.0in 0.0in 0.0in 0.0in,width=0.9\textwidth]{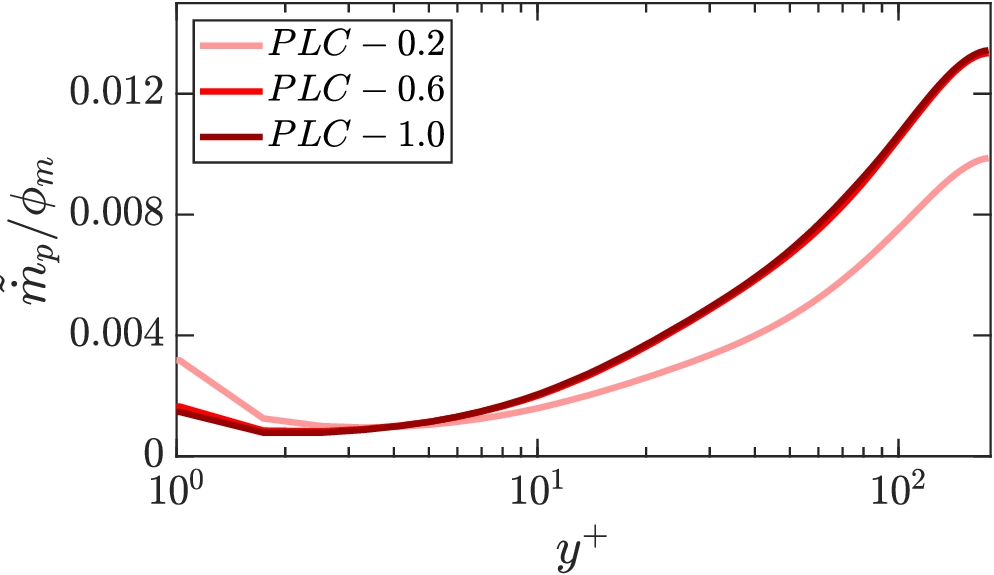}
                \end{minipage}
			\label{fig:m_dot_p_6}
		\end{subfigure}
		\begin{subfigure}{0.49\textwidth}
			\begin{minipage}[]{0.05\linewidth}
                    \subcaption{}
                \end{minipage}

                \begin{minipage}[]{1\linewidth}
                    \includegraphics[clip=true, trim = 0.0in 0.0in 0.0in 0.0in,width=0.9\textwidth]{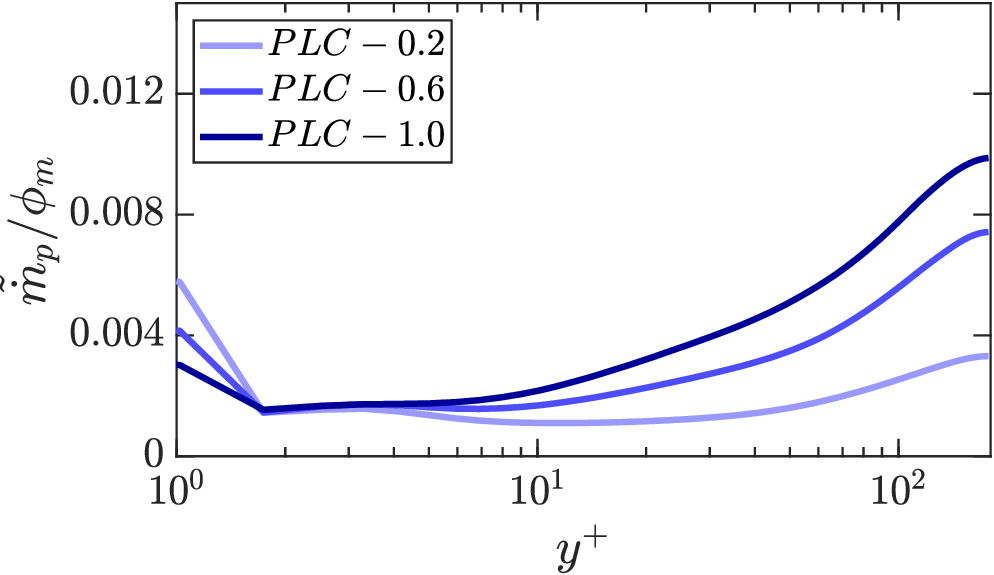}
                \end{minipage}
			\label{fig:m_dot_p_30}
		\end{subfigure}
        \caption{Variation of normalised particle average mass flow rate  ($\tilde{\dot{m}}_p = \overline{\dot{m}}_p/m_0$) per unit mass loading along the channel height for different PLC flows with (a) $St^+ = 6$ and (b) $St^+ = 30$. Darker curves correspond to higher particle mass loadings varying from $0.2$ to $1.0$.}
		\label{fig:m_dot_p}
	\end{figure}
    
    \subsubsection{\label{subsubsec:rss}Total Reynolds shear stress of the carrier flow}

    \quad The contribution of particles to the total RSS in the flow can be modelled by considering the streamwise momentum balance for a statistically steady flow. The fluid momentum balance in the streamwise direction gives
  \begin{equation}
	\frac{\partial (\rho (\overline{u}+u')(\overline{u}+u'))}{\partial x} + \frac{\partial (\rho v'(\overline{u}+u'))}{\partial y} = -\frac{\partial \overline{P}}{\partial x} - \frac{\partial P'}{\partial x} + \mu\frac{\partial^2 (\overline{u}+u')}{\partial y^2} - F_{Dx}
    \end{equation}
    For a fully developed channel flow, the mean pressure gradient is equal to the wall drag \citep{Pope_2000}:
     \begin{equation}
        \implies \frac{\partial (\rho v'\overline{u})}{\partial y} +  \frac{\partial (\rho u'v')}{\partial y} = \frac{\tau_w}{h} - \frac{\partial P'}{\partial x} + \mu\frac{\partial^2 \overline{u}}{\partial y^2} + \frac{\partial^2 u'}{\partial y^2} - F_{Dx}
    \end{equation}
    After averaging over time:
    \begin{equation}
        \implies \frac{\partial \overline{\rho u'v'}}{\partial y} = \frac{\tau_w}{h} + \mu\frac{\partial^2 \overline{u}}{\partial y^2} - \overline{F}_{Dx}
       \label{eq:rey_x_mom}
   \end{equation}

Similarly, the particle momentum balance in the streamwise direction gives
    \begin{equation}
	\frac{\partial \overline{\rho_p u_p'v_p'}}{\partial y} = \overline{F}_{Dx}
        \label{eq:rey_x_mom_p}
    \end{equation}

    Adding (\ref{eq:rey_x_mom}) to (\ref{eq:rey_x_mom_p}):
   \begin{equation}
        \frac{\partial \overline{\rho u'v'}}{\partial y} + \frac{\partial \overline{\rho_p u_p'v_p'}}{\partial y} = \frac{\tau_w}{h} + \mu\frac{\partial^2 \overline{u}}{\partial y^2}
    \end{equation}
Finally, integrating from $y$ to $h$:
    \begin{equation}
        - \overline{\rho u'v'} - \overline{\rho_p u_p'v_p'} = \tau_w\bigg(1-\frac{y}{h}\bigg) - \mu\frac{\partial \overline{u}}{\partial y}
    \end{equation}
    \begin{equation}
        \mu\frac{\partial \overline{u}}{\partial y} = \tau_w\bigg(1-\frac{y}{h}\bigg) - (-\overline{\rho u'v'} -\overline{\rho_p u_p'v_p'})
        \label{eq:rey_x_mom_total}
    \end{equation}
        
    \noindent where $\tau_w$, being a function of $Re_\tau$, is constant. While $-\overline{\rho u'v'}$ gives the fluid RSS, $-\overline{\rho_p u_p'v_p'}$ is the particle RSS. It can be concluded from (\ref{eq:rey_x_mom_total}) that at a particular distance from the wall, the RSS in both phases tends to reduce the viscous stress by reducing the mean streamwise velocity gradient. The lower the RSS, the greater the streamwise velocity gradient. The particles' effect on the RSS in both phases is shown in figure \ref{fig:rss_p}. The addition of particles reduces the fluid RSS, and the reduction increases monotonically with the increase in $\phi_m$. Furthermore, slightly higher values of the fluid RSS can be observed for $St^+ = 30$ cases than that for $St^+ = 6$ cases at the same $\phi_m$. This reduction in the fluid RSS can be attributable to the dissipation in fluid velocity fluctuations in the wall-normal direction as shown in figure \ref{fig:plc_fluc}. Moreover, \cite{fukagata2002contribution} analytically explained that the fluid RSS can be reduced by the blowing effect at the channel wall. In this context, blowing can be considered equivalent to the positive drag of particles on the fluid ($= -\overline{F}_{Dx}$, refer to figure \ref{fig:drag_force}) near the wall, which causes a reduction in the fluid RSS. 
    
    The particle RSS increases with an increase in $\phi_m$. From (\ref{eq:rey_x_mom_p}), it can be interpreted that the particle RSS is a depiction of the mean drag between the two phases. Assuming a very low particle Reynolds number, the mean of this interphase streamwise drag per unit volume can be formulated as
    \begin{equation}
        \overline{F}_{Dx} = \frac{18 \mu}{d_p^2} \overline{(u - u_p)\phi_v} = \frac{\rho_{pp}Re_\tau^2\mu}{h^2\rho}\frac{\overline{(u - u_p)\phi_v}}{St^+}
        \label{eq:mean_drag}
    \end{equation}

    $\frac{\rho_{pp}Re_\tau^2\mu}{h^2\rho}$ is a constant in all cases. Figure \ref{fig:drag_force} shows the variation of this mean drag force along the channel height. For the case of $St^+ = 30$ and $\phi_m = 1.0$, the plot using the present approach validates with that using the Lagrangian simulation by \cite{zhao2013interphasial}. Near the walls, the particle velocity is greater than that of the fluid, resulting in a negative drag. The higher particle velocity at the wall can be attributed to the slip and reflective nature of particle-wall interactions as opposed to the no-slip condition of the fluid phase. Away from the wall, the fluid eventually gains momentum under the effect of the pressure gradient and decreased wall effects, resulting in a higher velocity than that of the particle phase. This causes a positive mean drag. The effect of $\phi_m$ is clear from the figure; as $\phi_m$ increases, more particles drag the fluid, resulting in an enhancement in both the positive and negative drags. This can also be inferred from (\ref{eq:mean_drag}) that shows that the interphase drag directly depends on the particle volume fraction $\phi_v$. As stated in \textsection \ref{subsec:parclus}, $\phi_v$ at any location increases with $\phi_m$, thus increasing the drag. As a result, the particle RSS increases with an increase in $\phi_m$, as shown in figures \ref{fig:rss_p}(c,d).   
    \begin{figure}
		\begin{subfigure}{0.49\textwidth}
                \begin{minipage}[]{0.05\linewidth}
                    \subcaption{}
                \end{minipage}

                \begin{minipage}[]{1\linewidth}
                    \includegraphics[clip=true, trim = 0.0in 0.0in 0.0in 0.0in,width=0.9\textwidth]{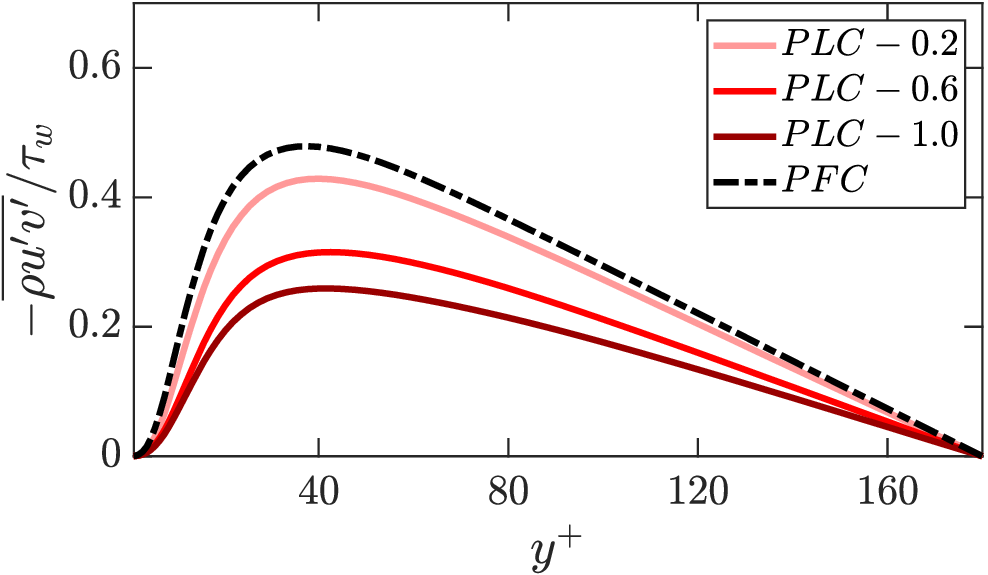}
                \end{minipage}
			\label{fig:rssf_6}
		\end{subfigure}
            \begin{subfigure}{0.49\textwidth}
			\begin{minipage}[]{0.05\linewidth}
                    \subcaption{}
                \end{minipage}

                \begin{minipage}[]{1\linewidth}
                    \includegraphics[clip=true, trim = 0.0in 0.0in 0.0in 0.0in,width=0.9\textwidth]{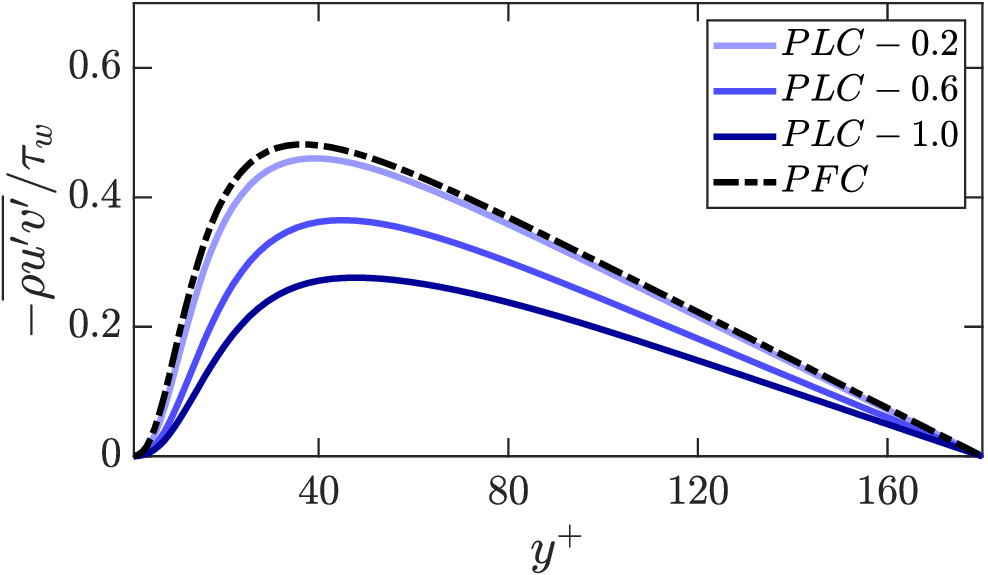}
                \end{minipage}
			\label{fig:rssf_30}
		\end{subfigure}
  
            \begin{subfigure}{0.49\textwidth}
			\begin{minipage}[]{0.05\linewidth}
                    \subcaption{}
                \end{minipage}

                \begin{minipage}[]{1\linewidth}
                    \includegraphics[clip=true, trim = 0.0in 0.0in 0.0in 0.0in,width=0.9\textwidth]{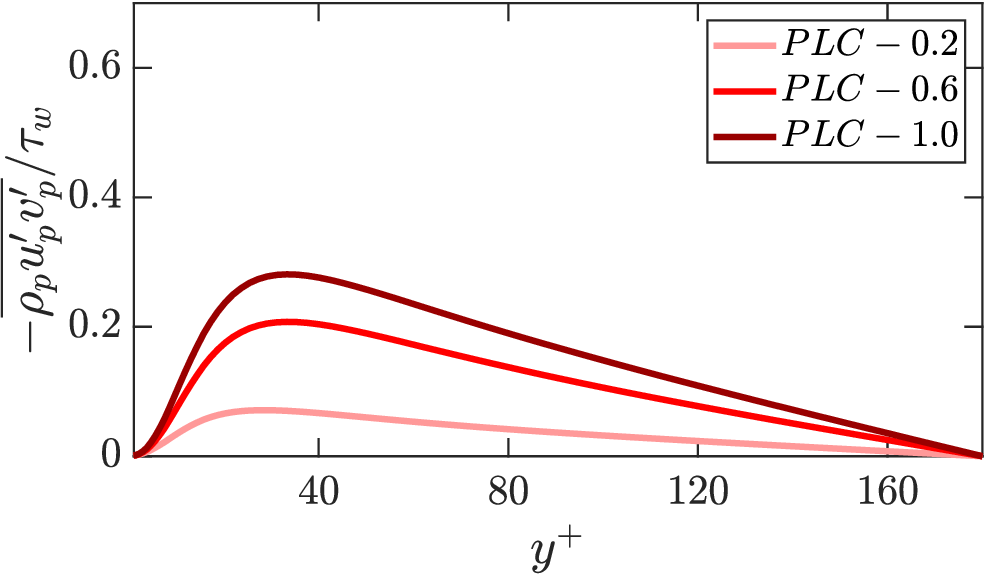}
                \end{minipage}
			\label{fig:rssp_6}
		\end{subfigure}
		\begin{subfigure}{0.49\textwidth}
			\begin{minipage}[]{0.05\linewidth}
                    \subcaption{}
                \end{minipage}

                \begin{minipage}[]{1\linewidth}
                    \includegraphics[clip=true, trim = 0.0in 0.0in 0.0in 0.0in,width=0.9\textwidth]{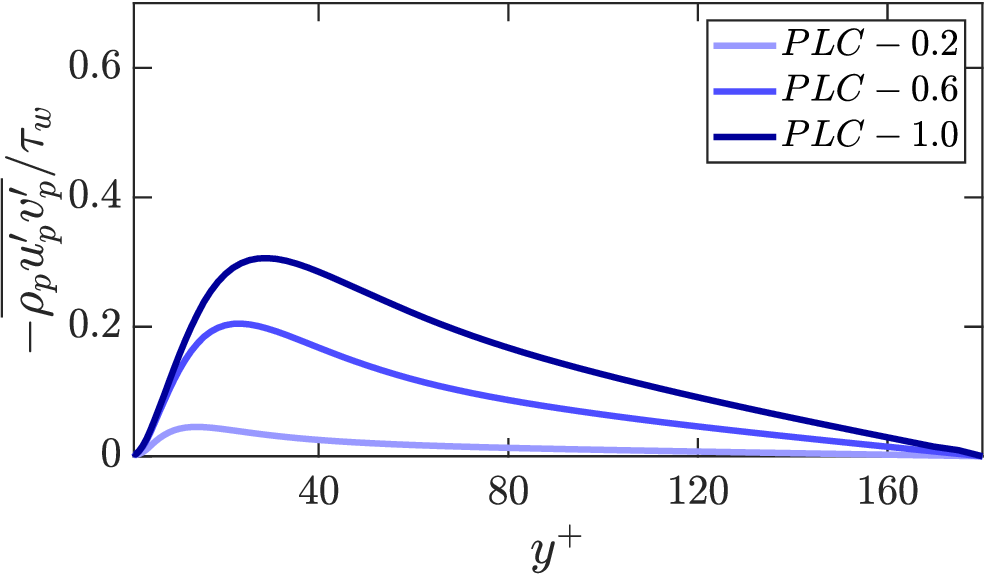}
                \end{minipage}
			\label{fig:rssp_30}
		\end{subfigure}
 
            \caption{Variation of different RSS along the channel height. The fluid RSS in the PFC flow and different PLC flows are shown in (a) for $St^+ = 6$ and (b) for $St^+ = 30$, while the particle RSS in different PLC flows are shown in (c) for $St^+ = 6$ and (d) for $St^+ = 30$. Darker curves correspond to higher particle mass loadings varying from $0.2$ to $1.0$.}
		\label{fig:rss_p}
	\end{figure}
        \begin{figure}
		\begin{subfigure}{0.49\textwidth}
			\begin{minipage}[]{0.05\linewidth}
                    \subcaption{}
                \end{minipage}

                \begin{minipage}[]{1\linewidth}
                    \includegraphics[clip=true, trim = 0.0in 0.0in 0.0in 0.0in,width=0.9\textwidth]{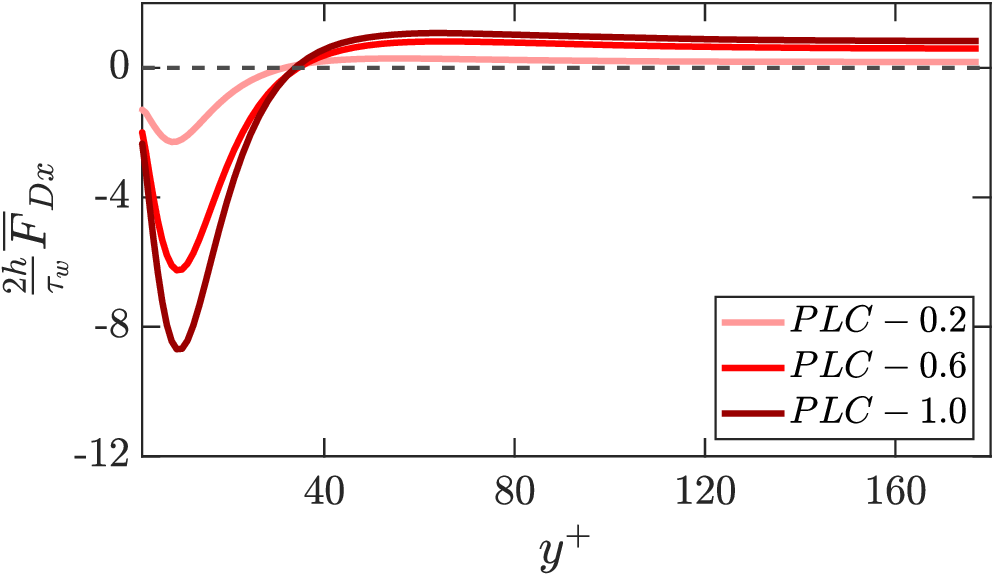}
                \end{minipage}
			\label{fig:fdp_p_6}
		\end{subfigure}
            \begin{subfigure}{0.49\textwidth}
			\begin{minipage}[]{0.05\linewidth}
                    \subcaption{}
                \end{minipage}

                \begin{minipage}[]{1\linewidth}
                    \includegraphics[clip=true, trim = 0.0in 0.0in 0.0in 0.0in,width=0.9\textwidth]{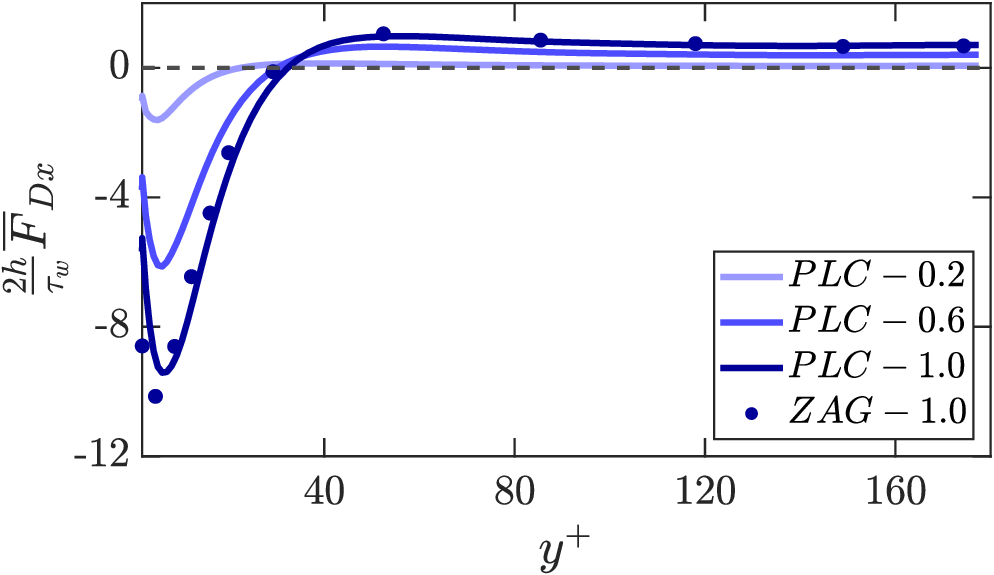}
                \end{minipage}
			\label{fig:fdp_p_30}
		\end{subfigure}
 
            \caption{Variation of mean of interphase drag per unit volume between the fluid phase and the particle phase along the channel height for different PLC flows with (a) $St^+ = 6$ and (b) $St^+ = 30$. The drag is normalised by $2h/\tau_w$. Darker curves correspond to higher particle mass loadings varying from $0.2$ to $1.0$. The dashed line separates the positive and negative drag. For the case of $St^+ = 30$ and $\phi_m = 1.0$, the results are compared to those obtained using the Lagrangian simulation by \cite{zhao2013interphasial} (ZAG; shown using symbols).}
		\label{fig:drag_force}
	\end{figure}
    
    The effect of particles' inertia on the interphase drag can be analysed by comparing the two subplots in figure \ref{fig:drag_force}. For clarity, this comparison at $\phi_m = 0.6$ is shown in figure \ref{fig:drag_force_comp}. Near the channel centre, the drag between the low-inertia particles ($St^+ = 6$) and the fluid is slightly more than that between the high-inertia particles ($St^+ = 30$) and the fluid. It is shown in figure \ref{fig:phi_p} that for the same $\phi_m$, low-inertia particles exhibit higher $\phi_v$ near the channel centre than high-inertia particles. Therefore, following (\ref{eq:mean_drag}), a higher positive interphase drag in the PLC flow with $St^+ = 6$ than in the PLC flow with $St^+ = 30$ can be concluded near the channel centre. Due to the higher drag between the fluid and the low-inertia particles, the fluid and particles have a greater likelihood of attaining equal velocities. Hence, the switching of the drag from positive to negative takes place at a closer distance from the centre than when $St^+ = 30$. Near the wall, although high-inertia particles exhibit a higher $\phi_v$, we do not observe a direct effect of $St^+$ on the drag magnitude. Similar trends are observed for other values of $\phi_m$.  Overall, the figure indicates that the effect of $St^+$ on the interphase drag is insignificant. As a result, the particle RSS is not significantly affected by the particle inertia, as can be inferred from figures  \ref{fig:rss_p}(c,d). Thus, it can be concluded that while the particle RSS increases with an increase in $\phi_m$ due to an increase in the interphase drag, $St^+$ has an insignificant effect on the particle RSS.
    \begin{figure}
        \centering
        \includegraphics[width=0.5\linewidth]{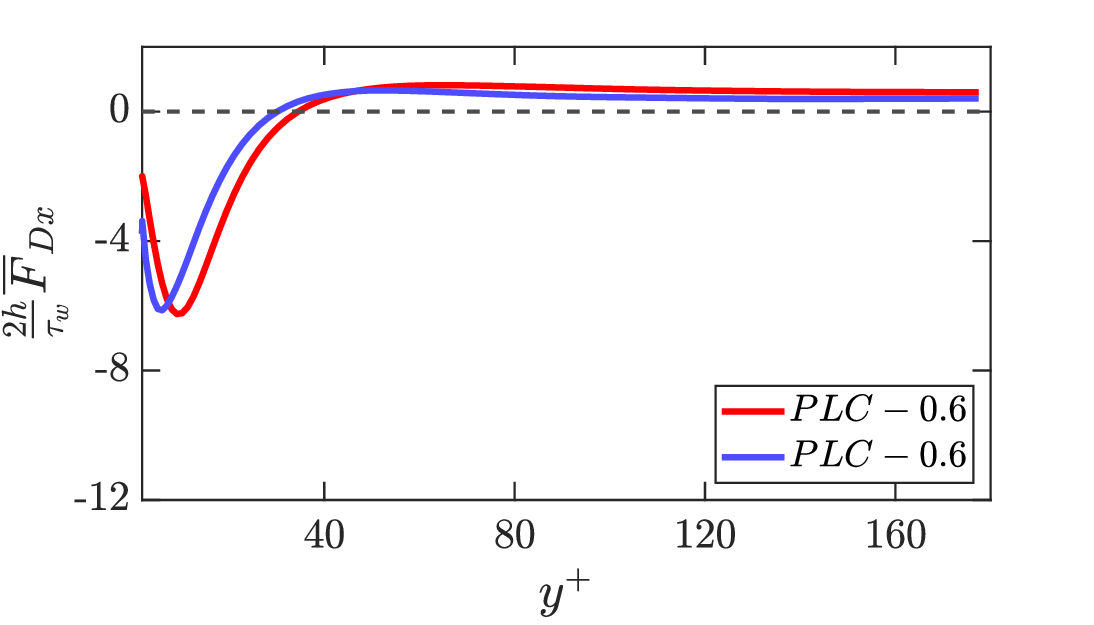}
        \caption{Comparison of fluid-particle interphase drag between PLC flows with $St+ = 6$ and $St^+ = 30$, at the same $\phi_m = 0.6$. The dashed line separates the positive and negative drag.}
        \label{fig:drag_force_comp}
    \end{figure}
    
    \begin{figure}
        \centering
         \begin{subfigure}{0.49\textwidth}
            \begin{minipage}[]{0.05\linewidth}
                \subcaption{}
            \end{minipage}
            
            \begin{minipage}[]{1\linewidth}
			\includegraphics[clip=true, trim = 0.0in 0.0in 0.0in 0.0in,width=0.9\textwidth]{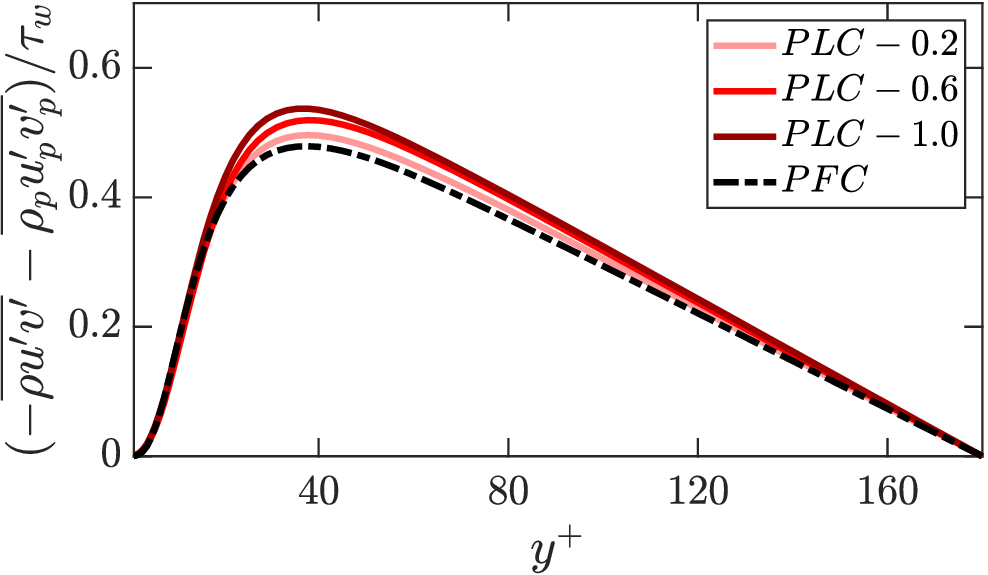}
            \end{minipage}
            \label{fig:rss_p_6}
		\end{subfigure}
		\begin{subfigure}{0.49\textwidth}
                \begin{minipage}[]{0.05\linewidth}
                \subcaption{}
                \end{minipage}

                \begin{minipage}[]{1\linewidth}
                    \includegraphics[clip=true, trim = 0.0in 0.0in 0.0in 0.0in,width=0.9\textwidth]{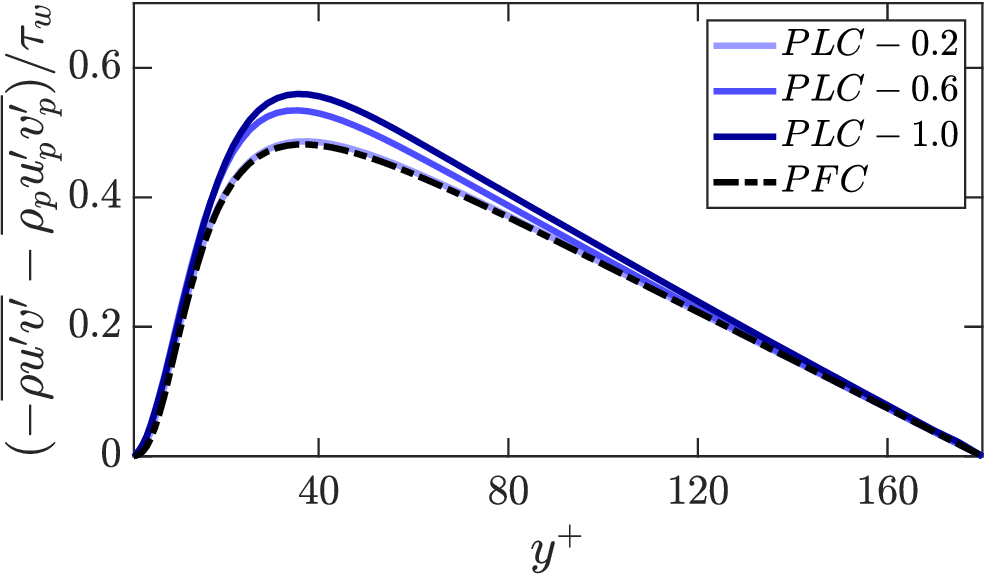}
                \end{minipage}
			\label{fig:rss_p_30}
		\end{subfigure}
        
         \caption{Variation of total RSS in the PFC flow and different PLC flows with (a) $St^+ = 6$ and (b) $St^+ = 30$. Darker curves correspond to higher particle mass loadings varying from $0.2$ to $1.0$.}
        \label{fig:total_rss}
    \end{figure}
    
    The effect of particles on the total RSS in both phases is represented in figure \ref{fig:total_rss}. There is no substantial effect of the particles on the total RSS in the region very close to the wall and to the channel centre, where the total RSS remains almost equal to the RSS in the PFC flow. In the region from $y^+ = 20$ to $y^+ = 120$, the total RSS increases to more than that of the PFC flow when the particles are added. For both $St^+$ cases, the total RSS increases with an increase in $\phi_m$.

    \subsubsection{\label{subsubsec:tur_mod}Particle streaks and turbulence modulation}
    \label{subsec:streaks}
    \quad With the effect of particles on fluid velocity fluctuations being anisotropic, we find that the addition of particles to the flow enhances the overall turbulent kinetic energy (TKE) of the fluid, as shown in figure \ref{fig:tke}. For $St^+ = 6$, the TKE increases when $\phi_m$ is increased from $0.2$ to $0.6$. However, we notice a decrease in the TKE upon a further increase of $\phi_m$ to $1.0$. This can be attributed to the saturation in $u_{rms}$ as described in figure \ref{fig:plc_fluc}. For $St^+ = 30$, the TKE increases upon increasing $\phi_m$. \cite{squires1994effect} studied particle-laden flows with Kolmogorov Stokes number in the range of 0.6-6.5, and $\phi_m$ ranging from 0 to 1.0. They found that for a significant $\phi_m$, turbulence structures can get distorted due to preferential clustering of particles, thus reducing the TKE. They showed that particles tend to increase the dissipation rate due to preferential clustering. Furthermore, \cite{squires1991preferential} reported that particles with low inertia exhibit a higher preferential concentration than particles with high inertia. For the same $\phi_m = 1.0$, it can be anticipated that particles in the flow with $St^+ = 6$ tend to cluster more than particles in the flow with $St^+ = 30$ (shown in table \ref{tab:parclus}). More clusters of particles interacting with the turbulence structures finally lead to a reduction in the TKE for the $St^+ = 6$ case.
        \begin{figure}

            \begin{subfigure}{0.49\textwidth}
			\begin{minipage}[]{0.05\linewidth}
                    \subcaption{}
                \end{minipage}

                \begin{minipage}[]{1\linewidth}
                    \includegraphics[clip=true, trim = 0.0in 0.0in 0.0in 0.0in,width=0.9\textwidth]{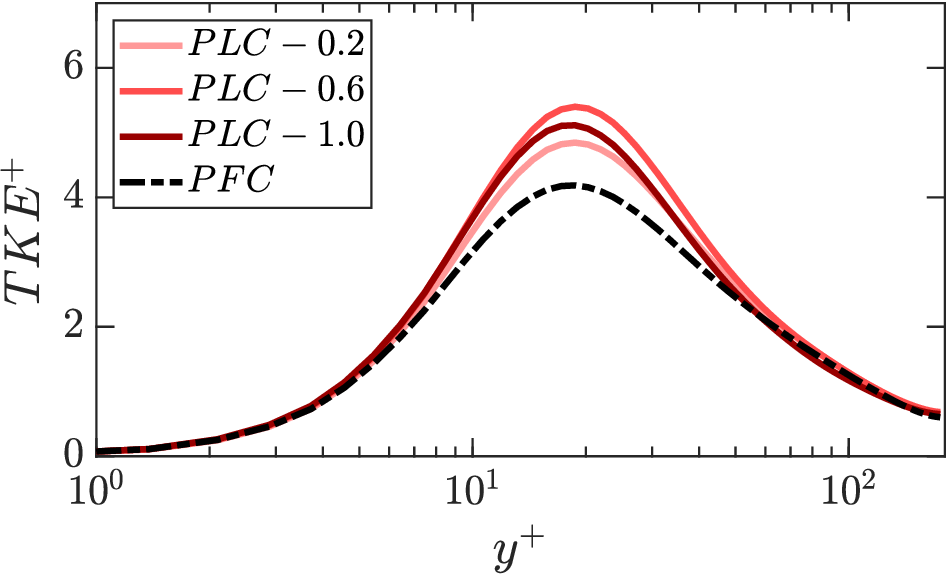}
                \end{minipage}
			\label{fig:tke_p_6}
		\end{subfigure}
		\begin{subfigure}{0.49\textwidth}
			\begin{minipage}[]{0.05\linewidth}
                    \subcaption{}
                \end{minipage}

                \begin{minipage}[]{1\linewidth}
                    \includegraphics[clip=true, trim = 0.0in 0.0in 0.0in 0.0in,width=0.9\textwidth]{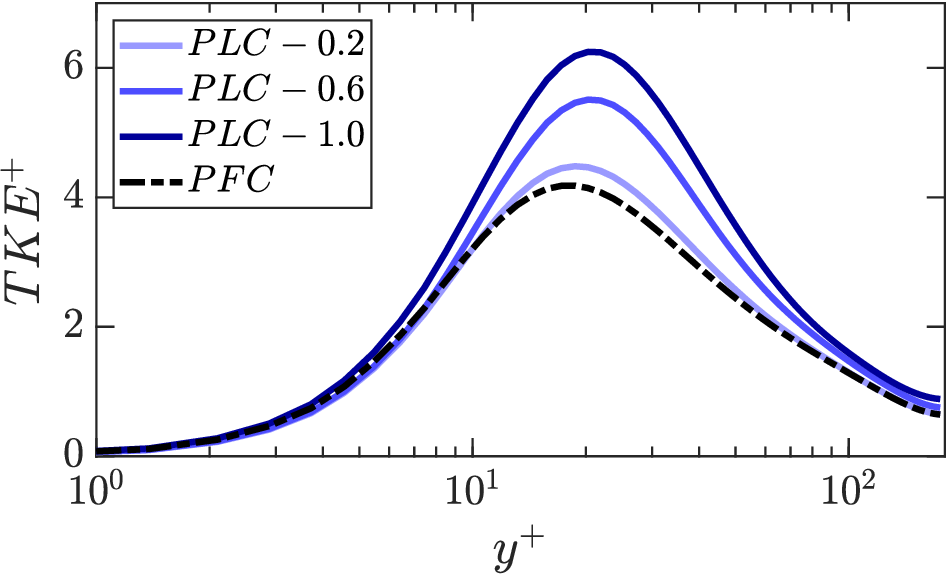}
                \end{minipage}
			\label{fig:tke_p_30}
		\end{subfigure}
  
		\caption{Variation of TKE along the channel height for the PFC flow and different PLC flows with (a) $St^+ = 6$ and (b) $St^+ = 30$. Darker curves correspond to higher particle mass loadings varying from $0.2$ to $1.0$.}
		\label{fig:tke}
	\end{figure}
    
        The presence of particle clusters in LSS can modify fluid turbulence by interacting with the turbulence structures \citep{squires1990particle}. \cite{zhou2020non} showed that adding particles to the fluid flow aligns and organises the fluid velocity streaks. For PLC flows with $St^+ = 30$, this is illustrated in figure \ref{fig:u_fluc_p_planey_30}, which shows the instantaneous contours of the fluid streamwise velocity fluctuations in the near-wall spanwise plane at $y^+ \approx 15$ at the same time instants as in figure \ref{fig:corr_p}. The contours are overlayed by red streaks of the particles $\phi_v$. The values of $\phi_v > 2.5\bar{\phi}_{vp}$ are shown to illustrate the regions of intense particle clustering. It can be noticed that the addition of particles aligns the turbulence streaks in the streamwise direction. Moreover, fewer streaks that are more organised and wider than those in the PFC flow are observed. 
    \begin{figure}
		\begin{subfigure}{0.49\textwidth}
                \begin{minipage}[]{0.05\linewidth}
                    \subcaption{}
                \end{minipage}

                \begin{minipage}[]{1\linewidth}
                    \includegraphics[clip=true, trim = 0.3in 0.0in 0.6in 0.1in,width=0.9\textwidth]{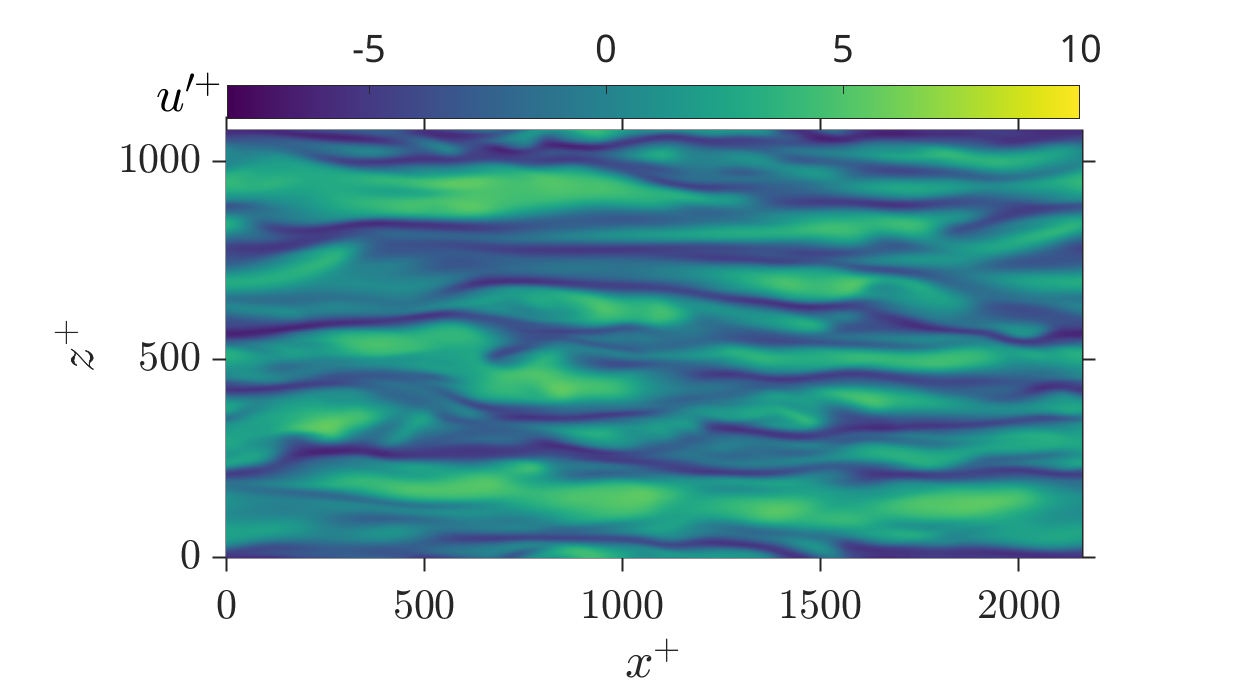}
                \end{minipage}
			\label{fig:u_fluc_p_planey_0_30}
		\end{subfigure}
		\begin{subfigure}{0.49\textwidth}
                \begin{minipage}[]{0.05\linewidth}
                    \subcaption{}
                \end{minipage}

                \begin{minipage}[]{1\linewidth}
                    \includegraphics[clip=true, trim = 0.3in 0.0in 0.6in 0.1in,width=0.9\textwidth]{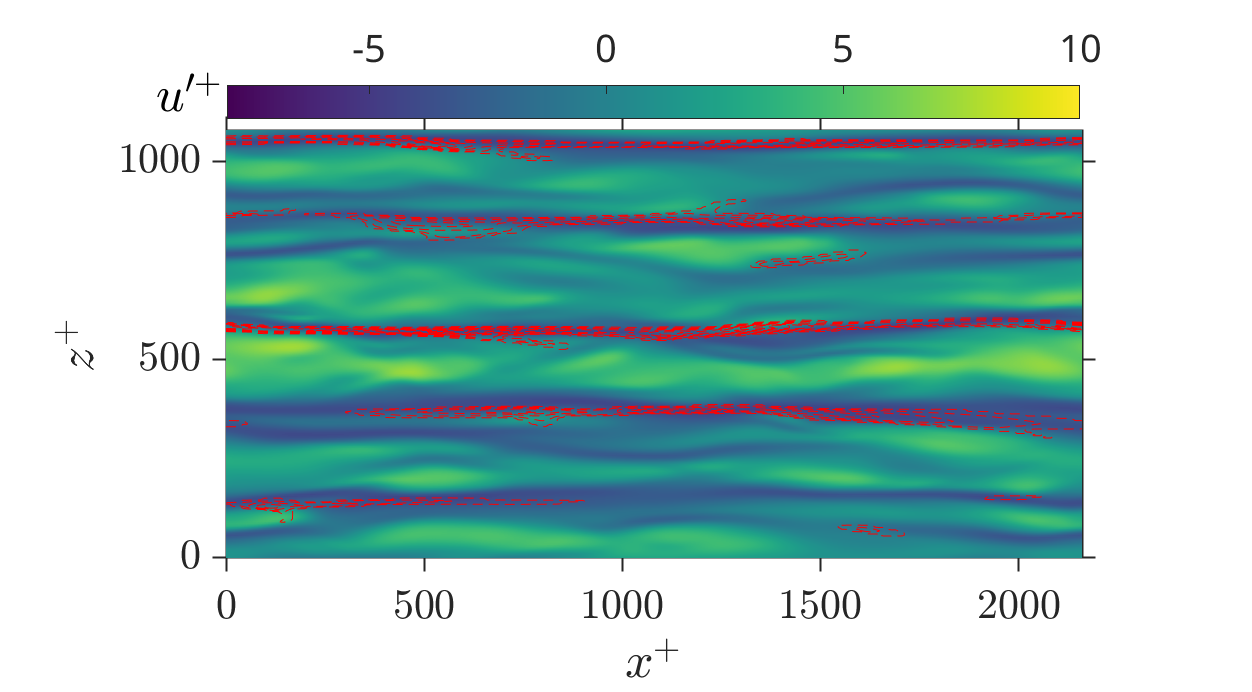}
                \end{minipage}
			\label{fig:u_fluc_p_planey_2_30}
		\end{subfigure}

            \begin{subfigure}{0.49\textwidth}
                \begin{minipage}[]{0.05\linewidth}
                    \subcaption{}
                \end{minipage}

                \begin{minipage}[]{1\linewidth}
                    \includegraphics[clip=true, trim = 0.3in 0.0in 0.6in 0.1in,width=0.9\textwidth]{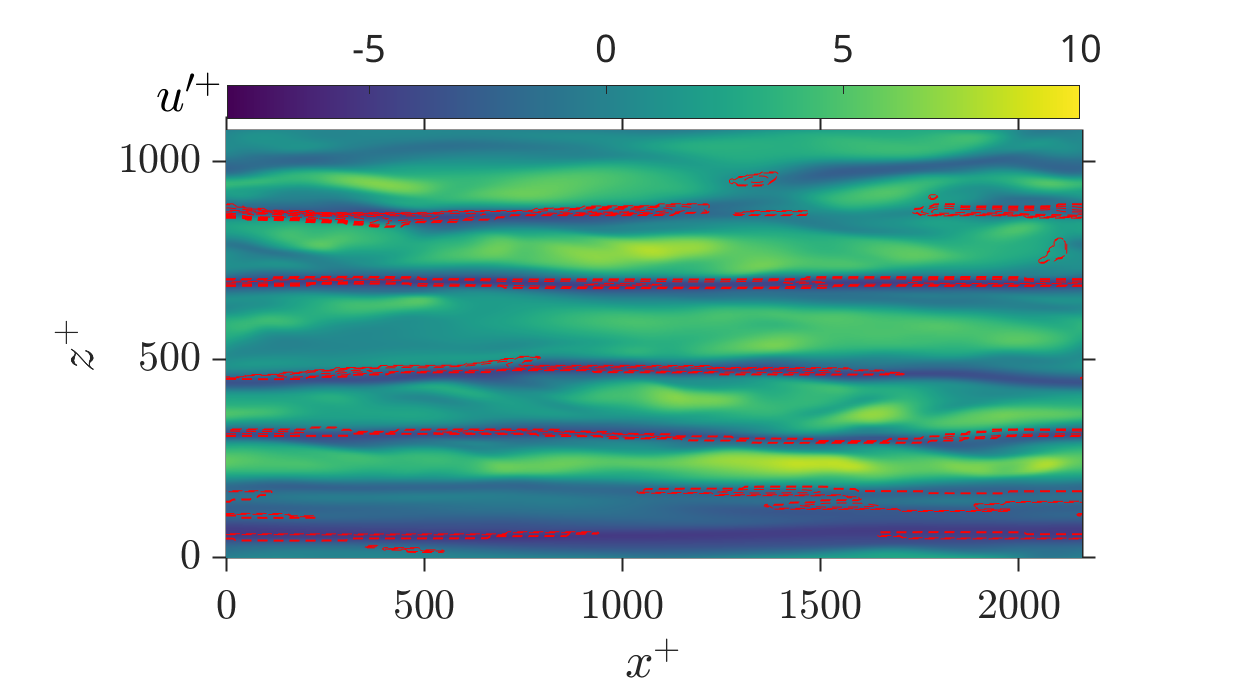}
                \end{minipage}
			\label{fig:u_fluc_p_planey_6_30}
		\end{subfigure}
		\begin{subfigure}{0.49\textwidth}
                \begin{minipage}[]{0.05\linewidth}
                    \subcaption{}
                \end{minipage}

                \begin{minipage}[]{1\linewidth}
                    \includegraphics[clip=true, trim = 0.3in 0.0in 0.6in 0.1in,width=0.9\textwidth]{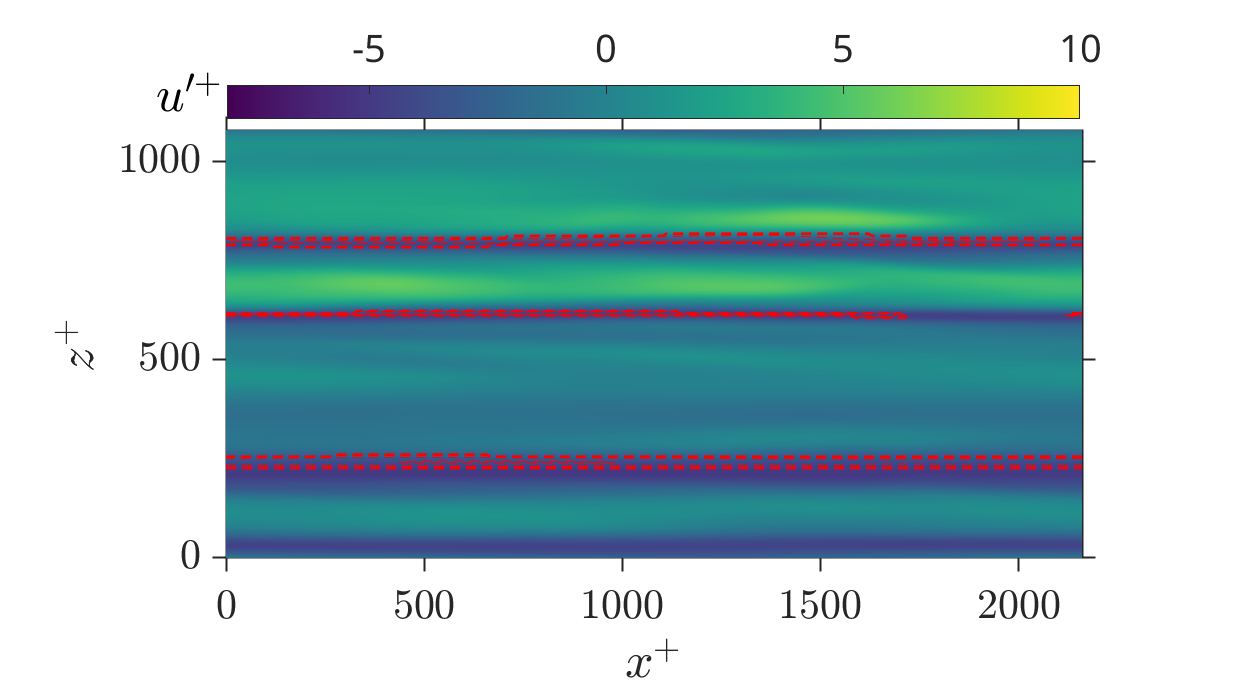}
                \end{minipage}
			\label{fig:u_fluc_p_planey_10_30}
		\end{subfigure}
  
		\caption{Contours showing the streaks of the fluid streamwise velocity fluctuations, overlayed by the red contours of relative particle volume fraction ($\phi_v > 2.5\bar{\phi}_{vp}$), in the spanwise plane at $y^+ \approx 10$ for $St^+ = 30$ at (a) $\phi_m = 0.0$, (b) $\phi_m = 0.2$, (c) $\phi_m = 0.6$, and (d) $\phi_m = 1.0$. Instantaneous data at the same time instants as in figure \ref{fig:corr_p} are analysed.}
		\label{fig:u_fluc_p_planey_30}
	\end{figure}
        \begin{figure}
		\begin{subfigure}{0.49\textwidth}
                \begin{minipage}[]{0.05\linewidth}
                    \subcaption{}
                \end{minipage}

                \begin{minipage}[]{1\linewidth}
                    \includegraphics[clip=true, trim = 0.0in 0.0in 0.6in 0.8in,width=0.9\textwidth]{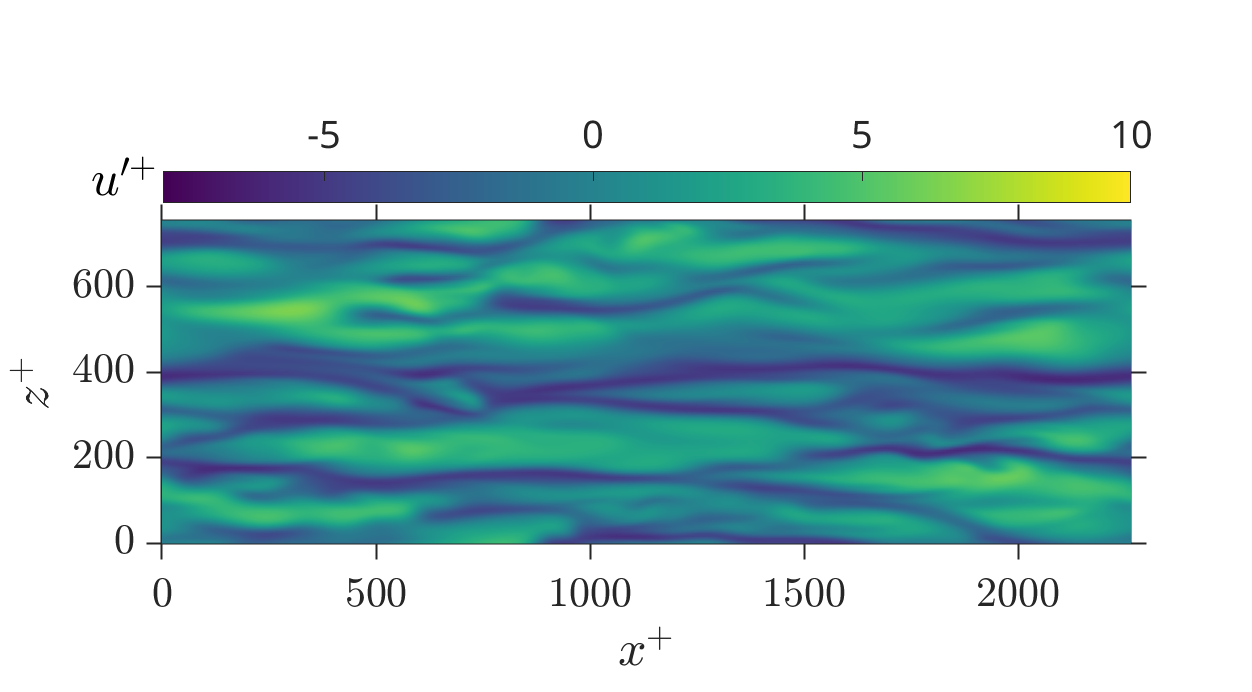}
                \end{minipage}
			\label{fig:u_fluc_p_planey_0_6}
		\end{subfigure}
		\begin{subfigure}{0.49\textwidth}
                \begin{minipage}[]{0.05\linewidth}
                    \subcaption{}
                \end{minipage}

                \begin{minipage}[]{1\linewidth}
                    \includegraphics[clip=true, trim = 0.0in 0.0in 0.6in 0.8in,width=0.9\textwidth]{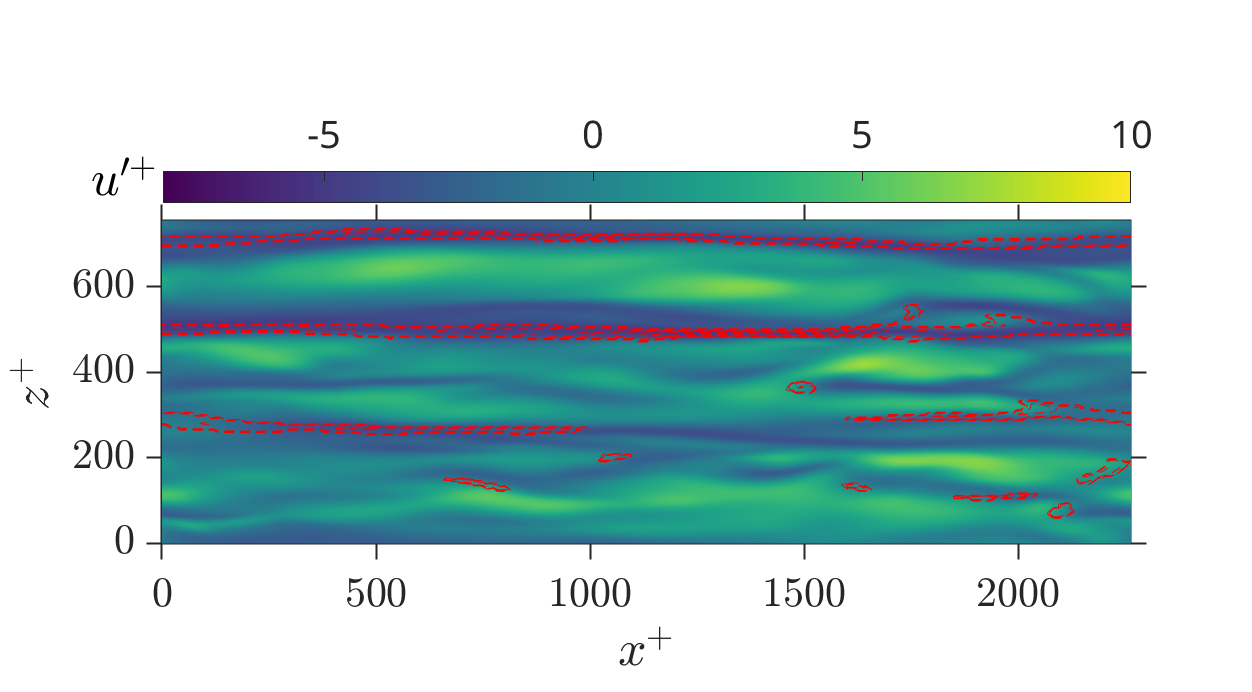}
                \end{minipage}
			\label{fig:u_fluc_p_planey_2_6}
		\end{subfigure}

            \begin{subfigure}{0.49\textwidth}
                \begin{minipage}[]{0.05\linewidth}
                    \subcaption{}
                \end{minipage}

                \begin{minipage}[]{1\linewidth}
                    \includegraphics[clip=true, trim = 0.0in 0.0in 0.6in 0.8in,width=0.9\textwidth]{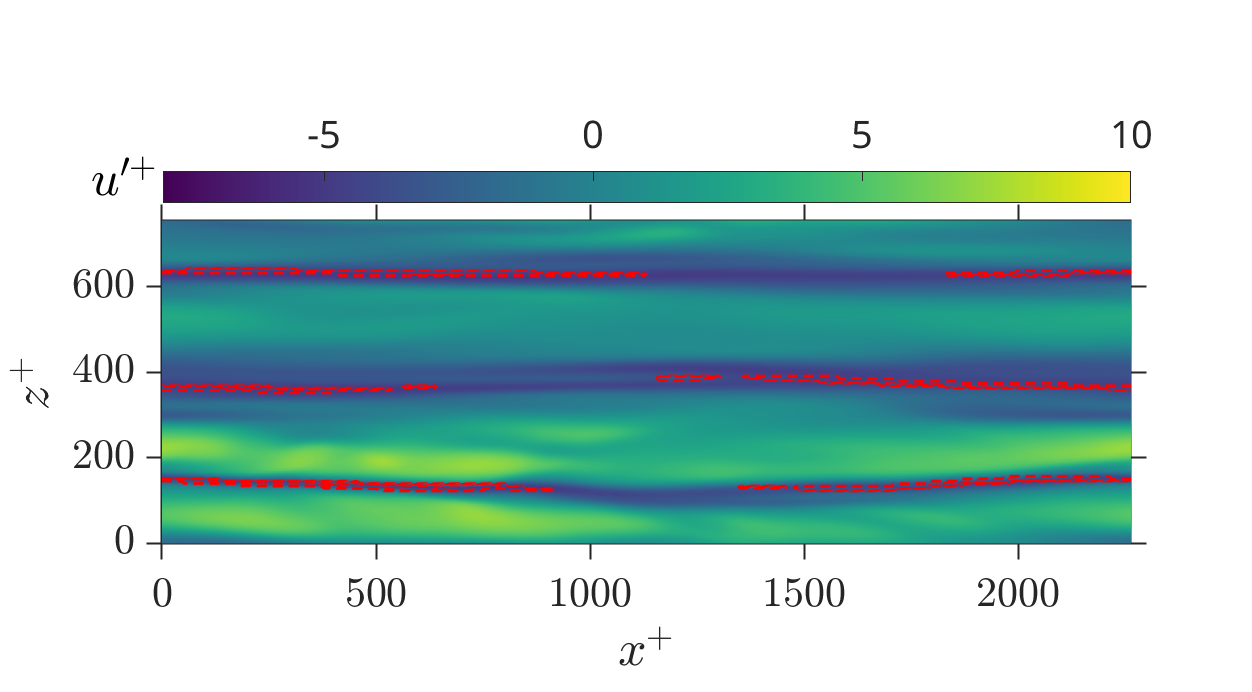}
                \end{minipage}
			\label{fig:u_fluc_p_planey_6_6}
		\end{subfigure}
		\begin{subfigure}{0.49\textwidth}
                \begin{minipage}[]{0.05\linewidth}
                    \subcaption{}
                \end{minipage}

                \begin{minipage}[]{1\linewidth}
                    \includegraphics[clip=true, trim = 0.0in 0.0in 0.6in 0.8in,width=0.9\textwidth]{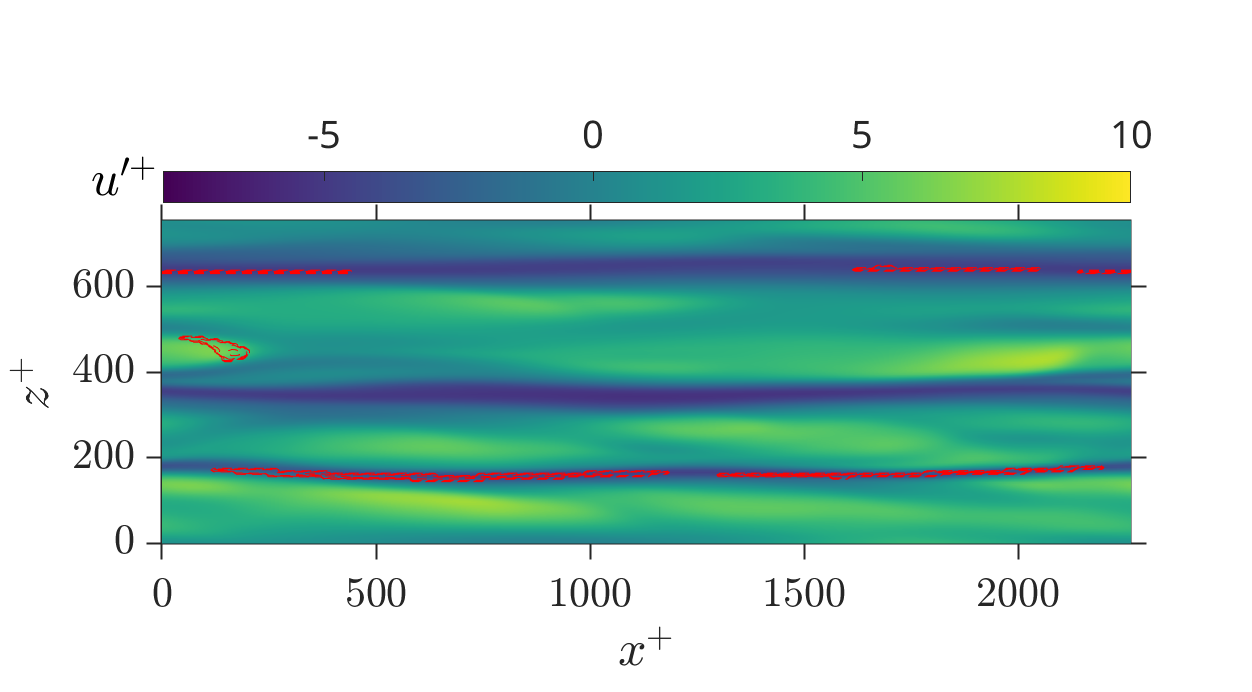}
                \end{minipage}
			\label{fig:u_fluc_p_planey_10_6}
		\end{subfigure}
  
		\caption{Contours showing the streaks of the fluid streamwise velocity fluctuations, overlayed by the red contours of relative particle volume fraction ($\phi_v > 2.5\bar{\phi}_{vp}$), in the spanwise plane at $y^+ \approx 15$ for $St^+ = 6$ at (a) $\phi_m = 0.0$, (b) $\phi_m = 0.2$, (c) $\phi_m = 0.6$, and (d) $\phi_m = 1.0$. Instantaneous data at the same time instants as in figure \ref{fig:corr_p} are analysed.}
		\label{fig:u_fluc_p_planey_6}
	\end{figure}
         The alignment and organisation of the turbulence streaks increase with an increase in $\phi_m$, while their number decreases. Fewer turbulence streaks in the flow should reduce turbulence. On the other hand, the strength of straightly aligned structures in the PLC flow increases with $\phi_m$, which may enhance turbulence \citep{zhou2020non}. The two competitive mechanisms modulate the TKE as shown in figure \ref{fig:tke}. The presence of high-density particle streaks in the LSS region can also be noticed in the figure. The spanwise spacing between the particle streaks aligns well with that between the LSS. Similar to the turbulent streaks, the particle streaks are also elongated. Thus, it can be concluded that particle streaks are interlinked to low-speed turbulence streaks and contribute to the modulation of flow turbulence.

    A similar analysis of PLC flows with $St^+ = 6$ is presented in figure \ref{fig:u_fluc_p_planey_6}. The effects of particles on the turbulence streaks are similar to those found in the case of PLC flows with $St^+ = 30$. Turbulent structures are more aligned at higher values of $\phi_m$, with the particle streaks being present in the region of LSS. Similar to $St^+ = 30$, it can be observed here that for the threshold value of $\phi_v > 2.5\bar{\phi}_{vp}$, the presence of particle streaks decreases as $\phi_m$ increases. At higher $\phi_m = 0.6$ and $1.0$, we observe the absence of particle streaks in some LSS regions. The presence of particle streaks is also lower than that in the PLC flows with $St^+ = 30$. However, referring to table \ref{tab:parclus}, it is possible that for a lower threshold value, more particle streaks will be present at higher $\phi_m$ and lower $St^+$. Thus, this observation on the effect of particle $St^+$ and $\phi_m$ on the particle streaks might change with the selection of the threshold value. Furthermore, the domain constraints may affect this analysis \citep{sardina2012wall} due to phase locking of largest turbulent structures. Nevertheless, it can be expected that even in the larger domains, these highly clustered streaks will appear in LSS in all the cases and will reorganise the streaks, and thus will modulate the turbulence.

\section{\label{sec:concl} Concluding remarks}
This study combines a quadrature three-order moment method with a compressible fluid solver to solve the particle field in a channel flow. We have demonstrated the capability and efficiency of this Eulerian approach in simulating dispersed particle-laden turbulent flows containing a very large number of particles in dilute regimes. This method offers a computationally less expensive alternative to the Lagrangian approach, which requires tracking each particle individually. The method can effectively simulate flows laden with very small Stokes numbers and volume fractions that can limit the simulation time step in Lagrangian approaches. The method is coupled with a compressible flow solver, adapted for incompressible flow with reduced numerical dissipation. Two-way coupling between the fluid and particles has been modelled using Stokes drag. The complete solver is developed in a GPU-accelerated framework.

We have demonstrated the quadrature method’s ability to predict the impact of particles on various fluid-particle interactions, such as mean flow and turbulence modulation, particle migration towards the wall, preferential clustering of particles, particle mass flow rate, and interphase drag between fluid and particles, which compares well with prior studies using the Lagrangian particle tracking approach. Furthermore, the effects of Stokes number and particle mass loading on these phenomena have been analysed in detail. The roles of particle migration to the wall and preferential clustering in modulating flow dynamics are also discussed.

Our findings reveal that introducing particles to the fluid increases the mean streamwise velocity, regardless of the Stokes number, though particle mass loading influences this increase. Up to a certain limit, adding more particles enhances the velocity, but beyond that, further particle addition reduces the mean velocity due to increased skin friction drag. Using the quadrature method, we show that particles amplify streamwise velocity fluctuations, while fluctuations in the other two directions are damped. However, at higher mass loadings, low-Stokes-number cases display a reduction in the streamwise fluctuations.

Particles tend to migrate towards the wall by the phenomenon of turbophoresis. This tendency of particles to accumulate in the low-turbulence region near the wall decreases with the increase in particle mass loading and increases with the increase in particle inertia. As a consequence of the conservation of the number of particles in the channel, the void density in the channel core decreases with the increase in particle mass loading and increases with the increase in particle inertia. Particles cluster preferentially along the low-speed turbulent streaks in the near-wall region. Although this preferential clustering is more significant for low-inertia particles, it is found to be mostly unaffected by the particle mass loading when the Stokes number is decreased. At a higher Stokes number, preferential clustering increases with an increase in particle mass loading. Particles' migration towards the wall affects their mass flow rate and the interphase drag across the channel height, while the preferential clustering of particles in the LSS modulates the turbulence.

The particle mass flow rate at the wall is not only non-zero due to slip and reflective wall conditions but also exhibits higher values than that in the near-wall regions. This is due to increased wall concentration of particles as a result of their migration towards the wall. In the near-wall region, the particle mass flow rate relative to the mass loading decreases as more particles are added to the flow. However, in the region away from the wall, mass loading has the opposite effect. This is a direct consequence of the particles' reduced tendency to migrate towards the wall with increased mass loadings. The particles' increased tendency to migrate towards the wall with an increase in Stokes number also affects the particle mass flow rate, which for higher Stokes numbers is higher in the near-wall region and lower in the region away from the wall.

By dampening the wall-normal velocity fluctuations, the addition of particles reduces the RSS in the fluid phase. However, by modelling the total stress balance in the streamwise direction, we realise the particle RSS, which increases with the addition of more particles. This stress is due to the drag between the two phases, which is negative near the wall and positive near the channel centre. This is due to particles having a higher velocity than the fluid phase in the near-wall region, which arises from the reflective wall boundary condition as opposed to the no-slip wall condition for the fluid. As the particle mass loading increases, the magnitudes of both positive and negative drags increase, and so does the particle RSS. However, the Stokes number does not have a significant effect on the interphase drag and, thus, on the particle RSS. Notably, the particle Stokes number affects the location where the drag changes sign from positive to negative, with lower Stokes numbers, shifting this point closer to the channel centre. This behaviour is explained by particle migration towards the wall due to turbophoresis. The low-inertia particles have higher concentrations near the channel centre than the high-inertia particles. Thus, the positive interphase drag is greater in flows seeded with low-inertia particles. Higher drag allows the particles and the fluid to reach the same velocity at a shorter distance from the channel centre, which leads to earlier switching of the drag profile when the Stokes number is small.

 Clusters of particles in the near-wall region interact with the turbulence streaks and align them along the streamwise direction. The number of streaks decreases with the addition of more particles, which can reduce turbulence. However, the streaks become more aligned and organised, thus enhancing their strength. The two competitive phenomena affect the overall flow turbulence. 

Thus, our results using the Eulerian quadrature-based moment method capture the direct effect of particle migration towards the wall on the particle mass flow rate. The results also reveal the physical mechanism between the particle accumulation towards the wall and the interphase drag between the two phases, which influence the particle RSS and, consequently, the viscous stress in the fluid. Finally, our method predicts the interplay between the particle preferential clustering and turbulence streaks near the wall, which modifies the flow turbulence. This study demonstrates how the Eulerian quadrature moment method approach can effectively resolve the proper dynamics of particle-laden turbulent flows and turbulence modulation by the dispersed particle phase. This Eulerian method may be extended for future research on a wide range of multiphase flow phenomena.

\appendix
\section{\label{app:matrix} Computational fluid dynamics matrix}
The appendix details the matrix of all fluid and particle variables used in the computational fluid dynamics simulation. Each row of the matrix corresponds to a particular equation in the transport model represented by (\ref{eq:transport_model}).
\begin{equation}
C_f = \begin{bmatrix}
        \rho \\
        \rho u \\
        \rho v \\
        \rho w \\
        \rho e \\
    \end{bmatrix} \hspace{4pt}
\vec{F}_f = \begin{bmatrix}
        \rho \vec{V} \\
        \rho u\vec{V} + P\hat{n}_x \\
        \rho v\vec{V} + P\hat{n}_y \\
        \rho w\vec{V} + P\hat{n}_z \\
        (\rho e + P)\vec{V} \\
      \end{bmatrix} \hspace{4pt}
\bar{\bar{S}}_v = \begin{bmatrix}
        0 \\
        \tau_{xx}\hat{n}_x + \tau_{xy}\hat{n}_y + \tau_{xz}\hat{n}_z \\
        \tau_{xy}\hat{n}_x + \tau_{yy}\hat{n}_y + \tau_{yz}\hat{n}_z \\
        \tau_{xz}\hat{n}_x + \tau_{yz}\hat{n}_y + \tau_{zz}\hat{n}_z \\
        q_x\hat{n}_x +  q_y\hat{n}_y + q_z\hat{n}_z \\
     \end{bmatrix} \hspace{4pt}
S_{c_f} = \begin{bmatrix}
        0 \\
        -F_{Dx} \\
        -F_{Dy} \\
        -F_{Dz} \\
        -E_D \\
      \end{bmatrix}
\end{equation}

\begin{equation}
C_p = \begin{bmatrix}
        M^0 \\[1pt]
        M^1_x \\[5pt]
        M^1_y \\[5pt]
        M^1_z \\[5pt]
        M^2_{xx} \\[5pt]
        M^2_{yy} \\[5pt]
        M^2_{zz} \\[5pt]
        Q \\
      \end{bmatrix} \hspace{4pt}
\vec{F}_p = \begin{bmatrix}
              M^1_x\hat{x} + M^1_y\hat{y} + M^1_z\hat{z} \\[1pt]
              M^2_{xx}\hat{x} + M^2_{xy}\hat{y}  + M^2_{xz}\hat{z} \\[5pt]
              M^2_{yx}\hat{x} + M^2_{yy}\hat{y} + M^2_{yz}\hat{z} \\[5pt]
              M^2_{zx}\hat{x} + M^2_{zy}\hat{y} + M^2_{zz}\hat{z} \\[5pt]
              M^3_{xxx}\hat{x} + M^3_{xxy}\hat{y} + M^3_{xxz}\hat{z} \\[5pt]
              M^3_{yyx}\hat{x} + M^3_{yyy}\hat{y} + M^3_{yyz}\hat{z} \\[5pt]
              M^3_{zzx}\hat{x} + M^3_{zzy}\hat{y} + M^3_{zzz}\hat{z} \\[5pt]
              R_x\hat{x} + R_y\hat{y} + R_z\hat{z} \\
            \end{bmatrix} \hspace{4pt}
\end{equation}
\begin{equation}
S_{c_p} = \begin{bmatrix}
        0 \\[1pt]
        \frac{\rho_{p1}}{m_p}F_{Dx1} + \frac{\rho_{p2}}{m_p}F_{Dx2} \\[5pt]
        \frac{\rho_{p1}}{m_p}F_{Dy1} + \frac{\rho_{p2}}{m_p}F_{Dy2} \\[5pt]
        \frac{\rho_{p1}}{m_p}F_{Dz1} + \frac{\rho_{p2}}{m_p}F_{Dz2} \\[5pt]
        2\frac{\rho_{p1}}{m_p}u_{p1}F_{Dx1} + 2\frac{\rho_{p2}}{m_p}u_{p2}F_{Dx2} \\[5pt]
        2\frac{\rho_{p1}}{m_p}v_{p1}F_{Dy1} + 2\frac{\rho_{p2}}{m_p}v_{p2}F_{Dy2} \\[5pt]
        2\frac{\rho_{p1}}{m_p}w_{p1}F_{Dz1} + 2\frac{\rho_{p2}}{m_p}w_{p2}F_{Dz2} \\[5pt]
        \sum\limits_{x,y,z}3(U_{p1}^2F_{D1})\frac{\rho_{p1}}{m_p} + 3\frac{\rho_{p2}}{m_p}\sum\limits_{x,y,z}(U_{p2}^2F_{D2}) \\[5pt]
      \end{bmatrix}
\end{equation}

\backsection[Funding]{This research was supported by the ISRAEL SCIENCE FOUNDATION (grant No. 1762/20).}

\backsection[Declaration of interests]{The authors report no conflict of interest.}

\backsection[Author ORCIDs]{Y. Dagan, https://orcid.org/0000-0002-7940-4002; Ajay Dhankarghare, https://orcid.org/0000-0001-8994-9463}

\bibliographystyle{jfm}
\bibliography{jfm, refTCFD,jfm0}

\end{document}